
\magnification = \magstep1
\vsize = 23.7 truecm  \hsize = 16 truecm
\baselineskip = 17pt

\centerline{\bf Influence of cubic and dipolar anisotropies on
the static and dynamic}
\centerline{\bf coexistence anomalies of the time-dependent
Ginzburg-Landau models}
\bigskip \bigskip
\centerline{U.C. T\"AUBER and F. SCHWABL}
\centerline{\it Institut f\"ur Theoretische Physik}
\centerline{\it Physik-Department der Technischen Universit\"at
M\"unchen}
\centerline{\it James-Franck-Str., D-W 8046 Garching, Germany}
\bigskip \bigskip

In isotropic systems below the transition temperature, the
massless Goldstone modes imply critical infrared singularities in
the statics and dynamics along the entire coexistence curve. We
examine the important question whether these coexistence
anomalies are of relevance also in more realistic systems
displaying anisotropies. By applying a generalized
renormalization scheme to the time-dependent Ginzburg-Landau
models, we treat two quite different but characteristic cases,
namely the influence of (i) weak cubic anisotropies, and (ii)
long-range dipolar interactions. In the presence of cubic terms,
the transverse excitations acquire a mass and thus one expects
the theory to approach an uncritical "Gaussian" regime in the
limit ${\vec q} \rightarrow 0$ and $\omega \rightarrow 0$.
Therefore, we first consider the one-component case in order to
show that our formalism also provides a consistent description of
the crossover into an asymptotically Gaussian theory. In the case
of (weak) cubic anisotropies, the fact that for $n < 4$ the
fluctuations tend to restore the $O(n)$-symmetry at the critical
point proves to be most important, since under these
circumstances coexistence-type singularities may be found in an
intermediate wavenumber and frequency range. The dipolar
interaction induces an anisotropy in momentum space, which does
not completely destroy the massless character of the transverse
fluctuations, but only effectively reduces the number of
Goldstone modes by one. Remarkably, similar to the isotropic case
the asymptotic theory can be treated exactly. For $n \geq 3$
we find coexistence anomalies governed by the isotropic power
laws. However, the amplitudes of the respective scaling functions
depend on the angle between order parameter and external
wavevector.
\bigskip
\leftline{\it PACS: 6460, 7540C, 7540G}
\vfill \eject

\centerline{\bf I. INTRODUCTION}
\bigskip
In this paper, we study the static and dynamic susceptibilities
of the $n$-component time-dependent Ginzburg-Landau models (also
called relaxational models A and B according to Hohenberg and
Halperin's classification [1]) below the transition temperature,
where in addition to an underlying $O(n)$-symmetry of the order
parameter field two characteristic anisotropies are taken into
account, namely (i) weak cubic anharmonic terms, and (ii)
long-range dipolar interactions.

Our considerations have the following background. In the ordered
phase of ideally isotropic systems, as a consequence of the
spontaneously broken symmetry there appear $n - 1$ massless
Goldstone modes which lead to infrared singularities in certain
correlation functions for all temperatures below the critical
temperature $T_c$. These so-called coexistence anomalies have
been studied on the basis of several exponentiation and
renormalization schemes for the static $\phi^4$-model [2-9] as
well as the dynamic relaxational models [5,10]. Lawrie's approach
[8] to take into account both critical and Goldstone fluctuations
within a suitable crossover theory, has recently been extended by
the present authors to the $O(n)$-symmetric time-dependent
Ginzburg-Landau models [11]. Generally, our calculations are
based on the path-integral formulation of dynamical perturbation
theory for Langevin-type equations of motion as developed by
Bausch, Janssen, Wagner [12] and De Dominicis [13], supplemented
by Amit and Goldschmidt's generalized renormalization procedure
especially designed for crossover phenomena [14]. To our opinion,
the formalism is rather transparent, the exact results in the
coexistence limit can be easily implemented, and the connections
to different approaches, e.g. the $1/n$-expansion [10,15,16], can
be very clearly demonstrated. One of the major additional
advantages is the possible applicability to even more complicated
models. Some of the central results for the dynamic
susceptibilities and correlation functions of the relaxational
models A and B are listed in Table I. We also remark that lately
a direct comparison with experimental findings of ultrasonic
attenuation measurements within the incommensurate phase of the
$A_2BX_4$-family crystals has been accomplished [17].

Having established a theory for ideally $O(n)$-symmetric systems,
it is quite natural a question to ask whether the striking
infrared anomalies found in the coexistence regime completely
disappear when (weak) anisotropies are introduced, or certain
remnants of the Goldstone excitations may still be experimentally
relevant. This issue is even more important as for many cases,
especially for structural and magnetic phase transformations with
an underlying discrete lattice symmetry, the assumption of an
isotropic order parameter field is certainly an idealization
which is only approximately justified. On the other hand, one
would expect a smooth transition of the physical properties from
an $O(n)$-symmetric model system to a real, anisotropic crystal.
Wallace [18] and Nelson [4] have noted that below but close to
$T_c$ even in weakly anisotropic sytems very strong fluctuations
induced by quasi-Goldstone modes should be prominent. The
dynamics in the high-frequency region of such cubic crystals in
the ordered phase was studied by Meissner, Menyh\'ard, and
Sz\'epfalusy [19]. Our focus, however, will be the crossover from
static and dynamic critical behavior into the hydrodynamic regime
of vanishing external wavenumbers and frequencies.

The outline of our paper is as follows. In the following Section
we briefly review the main results obtained for the
$O(n)$-symmetric relaxational models below the transition
temperature. The information presented here only serves for
comparison with the chapters below. For a thorough explanation
and discussion of the renormalization technique used, we refer
the interested reader to Ref.[11]. By specializing to the case $n
= 1$, we shall demonstrate in Section III that our procedure also
provides a consistent description of a crossover to an
asymptotically uncritical theory which can be described by a
Gaussian model. This constitutes an inevitable preliminary, as
the existence of cubic anisotropies in the free energy for the
order parameter field will eventually (i.e.: for ${\vec q}
\rightarrow 0$ and $\omega \rightarrow 0$) cause all the
fluctuations to cease. Under certain circumstances, however,
namely if in the vicinity of the critical point the
$O(n)$-symmetry is dynamically restored, coexistence-type
singularities may be found in an intermediate wavenumber and
frequency region, as we are going to point out in Section IV.
There, for $n < 4$ the crossover from the critical behavior to
the asymptotically Gaussian model via a Goldstone-mode dominated
regime will be discussed in detail, based on a one-loop
approximation. In Section V we shall turn to our second generic
example, namely long-range dipolar interactions introducing
anisotropies in momentum space. Our investigation of the
asymptotic theory, which remarkably can be treated exactly on
very much the same basis as the isotropic model, shows that
effectively the number of massless excitations is reduced by one.
Hence for $n \geq 3$ coexistence anomalies persist, characterized
by just the same power laws as summarized in Table I. However,
the amplitudes of the associated scaling functions are generally
lower, the reduction factors depending on the angle between the
direction of the static order parameter and the external
wavevector. At last we shall discuss and summarize our results.
In the Appendix we list the diagrams and the corresponding
analytical results for the two-point vertex functions up to
one-loop order for the time-dependent Ginzburg-Landau models
taking into account cubic terms.
\bigskip \bigskip

\centerline{\bf II. COEXISTENCE ANOMALIES IN THE ISOTROPIC}
\centerline{\bf RELAXATIONAL MODELS}
\bigskip
For convenience we start with a brief review of the static and
dynamic properties of the $O(n)$-symmetric time-dependent
Ginzburg-Landau models below the transition temperature. We shall
merely repeat the basic definitions of the models under
consideration and present the central results of the theory for
later comparison with the anisotropic cases. Further details,
especially concerning the dynamical perturbation expansion in the
ordered phase and the generalized renormalization scheme used to
describe the crossover behavior, may be found in our recent paper
[11].
\bigskip

\centerline{\bf A. Model equations and perturbation theory}
\medskip
We are interested in the statics and dynamics near a second-order
phase transition which phenomenologically can be described by an
$n$-component order parameter field $\phi_0^\alpha$, $\alpha = 1
, \ldots , n$, and the following $O(n)$-symmetric expansion of
the Ginzburg-Landau free-energy functional ("Hamiltonian") with
respect to $\phi_0^\alpha$,
$$H[\{ \phi_0^\alpha \}] = \int d^dx \left[ {r_0 \over 2} \,
\sum_{\alpha = 1}^n {\phi_0^\alpha({\vec x})}^2 + {1 \over 2} \,
\sum_{\alpha = 1}^n \Bigl[ {\vec \nabla} \phi_0^\alpha({\vec x})
\Bigr]^2 + {u_0 \over 4!} \, \left( \sum_{\alpha = 1}^n
{\phi_0^\alpha({\vec x})}^2 \right)^2 \right] \; .
\eqno(2.1a)$$
The parameter $r_0$ is proportional to the separation from the
mean-field transition temperature $T_c^0$, and hence can be
written as a sum of the fluctuation-induced $T_c$-shift $r_{0c}$
and the reduced temperature variable with respect to the critical
point $T_c$
$$r_0 = r_{0c} + {T - T_c \over T} \quad . \eqno(2.1b)$$
The positive coupling $u_0$ defines the strength of the isotropic
anharmonic term. In thermal equilibrium, the probability density
for a specific configuration $\phi_0^\alpha({\vec x})$ is
$$P[\{ \phi_0^\alpha \}] = {e^{- H[\{ \phi_0^\alpha \}]} \over
\int {\cal D}[\{ \phi_0^\alpha \}] \, e^{- H[\{ \phi_0^\alpha
\}]}} \quad . \eqno(2.2)$$

To constitute the dynamics of our model we assume that (i) the
only "slow" variable in the system is the order parameter itself,
i.e.: there is no conserved quantity besides perhaps
$\phi_0^\alpha$, and (ii) no reversible forces appear in the
generalized Langevin equation, which is the simplest conceivable
case. Therefore we have purely relaxational behavior where the
damping $\propto \lambda_0$ may be considered as originating from
the action of the "fast" degrees of freedom subsummed in the
fluctuating forces $\zeta^\alpha$. Thus our basic equation of
motion reads
$${\partial \over \partial t} \, \phi_0^\alpha({\vec x},t) =
- \lambda_0 \, (i \, {\vec \nabla})^a \, {\delta H[\{
\phi_0^\alpha \}] \over \delta \phi_0^\alpha({\vec x},t)} +
\zeta^\alpha({\vec x},t) \quad . \eqno(2.3)$$
Here two situations have to be distinguished, namely either there
is no conservation law for the order parameter field, and the
system simply relaxes into its equlibrium state after a
distortion [$a = 0$ in Eq.(2.3)], or $\phi_0^\alpha$ is a
conserved quantity itself and hence follows a diffusion equation
($a = 2$). These two cases are referred to as model A or model B,
respectively, according to the classification by Hohenberg and
Halperin [1].

If one takes the probability distribution of the stochastic
forces $\zeta^\alpha$ to be of a Gaussian form, one finds for the
first two moments
$$\eqalignno{\langle \zeta^\alpha({\vec x},t) \rangle &= 0
&(2.4a) \cr
\langle \zeta^\alpha({\vec x},t) \, \zeta^\beta({\vec x'},t')
\rangle &= 2 \, \lambda_0 \, (i \, {\vec \nabla})^a
\delta^{\alpha \beta} \, \delta({\vec x} - {\vec x'}) \, \delta(t
- t') \quad , &(2.4b) \cr}$$
which completes the definition of the time-dependent
Ginzburg-Landau models. The Einstein relation (2.4b) ensures that
the equilibrium probability density for a certain configuration
$\phi_0^\alpha({\vec x})$ is indeed given by Eq.(2.2).

Dynamical correlation functions are now calculated within the
field-theoretical approach to generalized Langevin equations as
developed by Bausch, Janssen, Wagner [12] and De Dominicis [13].
Starting from the assumed Gaussian distribution for the
stochastic forces, one eliminates the $\zeta^\alpha$ using the
equation of motion (2.3). After a Gaussian transformation and the
accompanying introduction of Martin-Siggia-Rose auxiliary fields
${\tilde \phi}_0^\alpha$, the probability density for the order
parameter fluctuations $\phi_0^\alpha$ finally reads
$$P[\{ \phi_0^\alpha \}] = \int {\cal D}[\{ i {\tilde
\phi}_0^\alpha \}] \, P[\{ {\tilde \phi}_0^\alpha \} , \{
\phi_0^\alpha \}] \propto \int {\cal D}[\{ i {\tilde
\phi}_0^\alpha \}] \, e^{J[\{ {\tilde \phi}_0^\alpha \} , \{
\phi_0^\alpha \}]} \quad , \eqno(2.5a)$$
where the statistical weight for a specific configuration is
determined by the so-called Janssen-De Dominicis functional $J$.
For the time-dependent Ginzburg-Landau models one finds
$$J[\{ {\tilde \phi}_0^\alpha \} , \{ \phi_0^\alpha \}] = \int \!
d^dx \! \int \! dt \sum_\alpha \left[ {\tilde \phi}_0^\alpha \,
\lambda_0 \, (i \, {\vec \nabla})^a \, {\tilde \phi}_0^\alpha -
{\tilde \phi}_0^\alpha \, \left( {\partial \phi_0^\alpha \over
\partial t} + \lambda_0 \, (i \, {\vec \nabla})^a \, {\delta H[\{
\phi_0^\alpha \}] \over \delta \phi_0^\alpha} \right) \right] \,
; \eqno(2.5b)$$
here we have omitted a term stemming from the functional
derivative of the transformation from the $\zeta^\alpha$ to the
$\phi_0^\alpha$, because it exactly cancels those contributions
to the perturbation series which violate causality [12].

Based on the path-integral formulation (2.5) for non-linear
Langevin dynamics, a perturbation expansion with respect to the
anharmonicity $\propto u_0$ and its diagrammatic representation
by propagators and vertices can be constructed in complete
analogy to the static case (see e.g. Ref.[20]). As usual, one
defines a generating functional for $N$-point Green functions and
cumulants by inserting source terms into (2.5b); furthermore,
after a Legendre transform one may also obtain the vertex
functions by appropriate functional derivatives. Details may be
found in Refs.[20,12,11].

In order to evaluate the dynamic (linear) response functions
within our formalism, we have to take into account the action of
an external field ${\tilde h}^\alpha$. This leads to an
additional term $- \sum_\alpha {\tilde h}^\alpha \,
\phi_0^\alpha$ in the Hamiltonian (2.1a), and thus to a modified
Janssen-De Dominicis functional [12]
$$J^{\tilde h}[\{ {\tilde \phi}_0^\alpha \} , \{ \phi_0^\alpha
\}] = J[\{ {\tilde \phi}_0^\alpha \} , \{ \phi_0^\alpha \}] +
\int \! d^dx \int \! dt \, \sum_\alpha {\tilde h}^\alpha \,
\lambda_0 \, (i \, {\vec \nabla})^a \, {\tilde \phi}_0^\alpha
\quad . \eqno(2.6)$$
Hence the dynamic susceptibility is given by the expression
$$\chi_0^{\alpha \beta}({\vec x},t;{\vec x'},t') = {\delta
\langle \phi_0^\alpha({\vec x},t) \rangle \over \delta
{\tilde h}^\beta({\vec x'},t')} \Bigg \vert_{{\tilde h}^\beta =
0} = \lambda_0 \, \langle \phi_0^\alpha({\vec x},t) \, (i \,
{\vec \nabla})^a \, {\tilde \phi}_0^\beta({\vec x'},t') \rangle
\quad , \eqno(2.7a)$$
and its Fourier transform is intimately connected with the
so-called response propagator
$$\chi_0^{\alpha \beta}({\vec q},\omega) = \lambda_0 \, q^a \,
G_{0 \, {\tilde \phi}^\alpha \phi^\beta}({\vec q},\omega) \quad .
\eqno(2.7b)$$
The dynamical correlation function can either be calculated
directly or derived from the following (classical)
fluctuation-dissipation theorem [12]
$$G_{0 \, \phi^\alpha \phi^\beta}({\vec q},\omega) = {2 \,
\lambda_0 \, q^a \, \over \omega} \, {\rm Im} \, G_{0 \, {\tilde
\phi}^\alpha \phi^\beta}({\vec q},\omega) \quad . \eqno(2.8)$$
\vfill \eject

\centerline{\bf B. Renormalization in the ordered phase and
coexistence anomalies}
\medskip
Below the critical temperature $T_c$ a spontaneous order
parameter ${\bar \phi}_0$ appears, which we assume to point in
the $n$-th direction of the order parameter space. For
convenience we define new transverse ($\alpha = 1, \ldots , n -
1$) and longitudinal fields
$${{\tilde \phi}_0^\alpha \choose {\tilde \phi}_0^n} = {{\tilde
\pi}_0^\alpha \choose {\tilde \sigma}_0} \quad , \qquad
{\phi_0^\alpha \choose \phi_0^n} = {\pi_0^\alpha \choose \sigma_0
+ {\bar \phi}_0} \eqno(2.9a)$$
with vanishing thermal average
$$\langle \pi_0^\alpha \rangle = \langle \sigma_0 \rangle = 0
\quad . \eqno(2.9b)$$
In addition we shall use the parametrization
$${\bar \phi}_0 = \sqrt{3 \over u_0} \, m_0 \quad , \eqno(2.10)$$
where the quantity $m_0$ has the dimension of a mass or inverse
length. By evaluating the condition (2.9b) for the longitudinal
field one arrives at the equation of state
$$r_0 + {m_0^2 \over 2} = A + {\cal O}(u_0^2) \quad ,
\eqno(2.11)$$
which allows the elimination of the temperature variable $r_0$ in
favor of the parameter $m_0$ [11,8]; to express the perturbation
expansion in terms of $m_0$ has the considerable advantage that
the $T_c$-shift is already properly included (compare the
discussion in Ref.[21]).

{}From a general dynamic Ward-Takahashi identity [11] the following
exact relation for the transverse two-point vertex function can
be derived
$${\bar \phi}_0 \, \Gamma_{0 \, {\tilde \pi} \pi}({\vec q} = 0,
\omega = 0) = {\tilde h}^n \quad , \eqno(2.12)$$
implying the massless character of the $n - 1$ transverse
excitations for vanishing external field ${\tilde h}^n
\rightarrow 0$. These Goldstone modes induce infrared
singularities that are not restricted to the vicinity of the
critical point, but persist in the entire ordered phase.

Very remarkably, however, in contrast to the critical region
($m_0 \approx 0$) the coexistence regime $T < T_c$, ${\vec q}
\rightarrow 0$, $\omega \rightarrow 0$, ${\tilde h}^n \rightarrow
0$ allows for an exact treatment [8,15,11]. Heuristically, the
above limits may be replaced by fixed wavenumbers and frequencies
and $m_0 \rightarrow \infty$ instead. The renormalization group
analysis a posteriori confirms the assertion that the infrared
behavior is indeed characterized by a diverging longitudinal mass
parameter $m$. By introducing new longitudinal fields
$$\varphi_0({\vec q},\omega) = m_0 \, \sigma_0({\vec q},\omega) +
{\sqrt{3 \, u_0} \over 6} \int_{q'} \! \int_{\omega'}
\sum_{\alpha = 1}^{n-1} \pi_0^\alpha({\vec q}',\omega') \,
\pi_0^\alpha({\vec q} - {\vec q}',\omega - \omega') + \sqrt{3
\over u_0} \, A \, \delta({\vec q}) \, \delta(\omega)
\eqno(2.13a)$$
$${\tilde \varphi}_0({\vec q},\omega) = m_0 {\tilde
\sigma}_0({\vec q},\omega) + {\sqrt{3 \, u_0} \over 3} \int_{q'}
\! \int_{\omega'} {{q'}^a \over q^a} \, \sum_{\alpha = 1}^{n-1}
{\tilde \pi}_0^\alpha({\vec q}',\omega') \, \pi_0^\alpha({\vec q}
- {\vec q}',\omega - \omega') \eqno(2.13b)$$
(provided ${\vec q} \not= 0$ in the case of model B) and
performing the limit $m_0 \rightarrow \infty$ --- under the
assumption $u_0 / m_0^2 \rightarrow 0$ --- one arrives at a
purely harmonic asymptotic functional. (Here we have introduced
the short-hand notation $\int_q \int_\omega \ldots \; = {1 \over
(2 \, \pi)^{d + 1}} \, \int d^dq \int d\omega \ldots \;$.) In the
coexistence regime merely transverse fluctuations exist, and the
corresponding Goldstone propagators are identical to those from
mean-field theory, as has been anticipated in earlier work
[2,3,5]. On the other hand, by using (2.13) one finds for the
(original) longitudinal correlation functions in the asymptotic
limit [11] exact results (see Eq.(3.18) of Ref.[11]) that
correspond to (i) the one-loop expressions for the two-point
cumulants, and (ii) the leading order of the $1/n$-expansion for
the two-point vertex functions [15,16], which can be summed up
via a geometric series (see Fig.2 of Ref.[11]). In the asymptotic
model the ultraviolet divergences can be removed by introducing
renormalized counterparts $m$ and $u$ for the order parameter and
coupling constant, respectively, where $m \propto u$ with a
finite proportionality factor. Within the renormalized theory,
the remaining infrared divergences are transcribed to an
anomalous dimension of the mass $m$ which eventually yields the
correct behavior for low wavenumbers and frequencies.

Following Lawrie's work [8], we apply Amit and Goldschmidt's
generalized minimal subtraction procedure [14] in order to
describe the entire crossover from the critical behavior to the
Goldstone regime. However, as we do not want to weaken the exact
statements valid in the coexistence limit by introducing any
further approximations, we refrain from $\epsilon$-expansion
[11], where
$$0 \leq \epsilon = 4 - d < 2 \eqno(2.14)$$
is the difference to the upper critical dimension of the
$\phi^4$-model. Following the arguments of Schloms and Dohm [21],
quite generally the extraction of the correct infrared behavior
via the analysis of the ultraviolet divergences appearing as
$\epsilon$-poles within the dimensional regularization scheme
does not require an expansion with respect to the distance from
the upper critical dimension. The latter is needed if one wants
to define a "small" expansion parameter in the perturbation
series which can be avoided by a suitable resummation procedure.
Our case is even simpler, though, for the theory asymptotically
reduces to the one-loop expressions.

Neglecting the Fisher corrections connected with the static
exponent $\eta$ we thus use a one-loop approximation for the
entire wavenumber and frequency region. On this level we can
drop any field renormalizations; furthermore, because the
$Z$-factor for the Onsager coefficient as a consequence of the
fluctuation-dissipation theorem (2.8) can be expressed in terms
of $Z_\phi$ and $Z_{\tilde \phi}$ [11,12] (see also Eq.(5.18)
below), one finds $Z_\lambda = 1 + {\cal O}(u_0^2)$ for the
relaxational models. Hence the only renormalized quantities we
have to introduce are
$$\eqalignno{m^2 &= Z_m^{-1} \, m_0^2 \, \mu^{-2} &(2.15a)\cr
u &= Z_u^{-1} \, u_0 \, A_d \, \mu^{- \epsilon} \quad ,
&(2.15b)\cr}$$
where for convenience we have explicitly separated the naive
dimensions and the geometric factor [21]
$$A_d = S_d \, \Gamma \! \left( 3 - {d \over 2} \right) \, \Gamma
\! \left( {d \over 2} - 1 \right) = {\Gamma (3 - {d \over 2})
\over 2^{d-2} \, \pi^{d \over 2} \, (d-2)} \quad . \eqno(2.16)$$

{}From the explicit one-loop results for the $Z$-factors defined in
(2.15) we proceed to calculate Wilson's flow functions
$$\eqalignno{\zeta_m(u,m) &= \mu \, {\partial \over \partial \mu}
\Big \vert_0 \ln \;  {m^2 \over m_0^2} = - 2 + {n - 1 \over 6} \,
u + {3 \over 2} \, {u \over (1 + m^2)^{1 + \epsilon / 2}}
&(2.17a)\cr
\beta_u(u,m) &= \mu \, {\partial \over \partial \mu} \Big \vert_0
\, u = u \, \left[ - \epsilon + {n - 1 \over 6} \, u + {3 \over
2} \, {u \over (1 + m^2)^{1 + \epsilon / 2}} \right] \quad .
&(2.17b)\cr}$$
They enter the renormalization group equations for the cumulants,
which are solved via the introduction of the characteristics
$\mu(\ell) = \mu \, \ell$ with the result
$$G^c_{{\tilde N} N}(\mu , \lambda , m , u ,\{ {\vec q} \},\{
\omega \}) = G^c_{{\tilde N} N}(\mu \ell , \lambda , m(\ell) ,
u(\ell) , \{ {\vec q} / \mu \ell \},\{ \omega / (\mu \ell)^2 \})
\quad , \eqno(2.18)$$
where the flow-dependent parameters are to be determined as
solutions of the coupled first-order differential equations
$$\eqalignno{\ell \, {d m(\ell) \over d \ell} &= {1 \over 2} \,
m(\ell) \, \zeta_m(\ell) &(2.19a)\cr
\ell \, {d u(\ell) \over d \ell} &= \beta_u(\ell) &(2.19b)\cr}$$
with the initial conditions $m(1) = m$, $u(1) = u$.

The fixed points of the renormalization group characterizing the
asymptotic behavior are given by the zeros of the
$\beta$-function (2.19b). At the critical point ($m = 0$) the
Heisenberg fixed point
$$u_H^* = {6 \, \epsilon \over n + 8} \eqno(2.20a)$$
becomes stable, and the associated anomalous dimension for the
mass parameter is
$${\zeta_m^*}_H = - 2 + \epsilon \quad . \eqno(2.20b)$$
On the contrary, in the limit $m \rightarrow \infty$ Lawrie's
coexistence fixed point [8] is approached
$$\eqalignno{u_C^* &= {6 \, \epsilon \over n - 1} &(2.21a)\cr
{\zeta_m^*}_C &= - 2 + \epsilon \quad ; &(2.21b)\cr}$$
hence indeed $m(\ell)^2 \propto \ell^{- 2 + \epsilon} \rightarrow
\infty$ for $\ell \rightarrow 0$ in the range $0 \leq \epsilon <
2$, and the above treatment of the asymptotic model becomes
meaningful. For $d > 4$, the Gaussian fixed point is stable
and the coexistence anomalies (as well as any non-trivial
critical behavior) disappear. At the critical dimension $d_c = 4$
there will be logarithmic corrections. On approaching $d
\rightarrow 2$, $m(\ell)$ no longer diverges and the assumption
of a homogeneous long-range order parameter turns out to be
inconsistent, which is to be expected according to the theorem by
Mermin, Wagner [22], and Hohenberg [23].

In Ref.[11] approximate analytical as well as numerical solutions
of the flow equations (2.19) are discussed in detail (see Fig.5
for $y = 0$). Here we just quote that because of $Z_m = Z_u$ the
ratio
$${m(\ell)^2 \over u(\ell)} \, \ell^{2 - \epsilon} = {m(1)^2
\over u(1)} \eqno(2.22)$$
is invariant under the renormalization group, and the important
property that for small initial values $m(1) < 0.1$, the flows
coincide when expressed in terms of the scaling variable
$$x = {\ell \over m(1)^{2 / (2 - \epsilon)}} \quad ,
\eqno(2.23)$$
which turns out to be the origin for the general scaling laws
obeyed by the static and dynamic susceptibilities. In order to
determine the scaling functions we finally have to use a matching
condition fixing at least one of the arguments on the right-hand
side of (2.18b) at a sufficiently large value such that a
conventional perturbation approach is applicable. In this context
a very suitable choice will be
$$\ell^2 = \bigg \vert {q^2 \over \mu^2} - { i \, \omega \over
\lambda \, \mu^2 \, (q / \mu)^a} \bigg \vert \quad ,
\eqno(2.24)$$
i.e.: $\ell^2 \approx \mid \chi_T^{-1}({\vec q},\omega) \mid$ is
precisely identified with the distance to the critical surface of
the Goldstone singularities.

Taking advantage of the scaling property (2.23) and inserting
(2.24) into the perturbational expressions for the dynamic
susceptibilities, one finds that the response functions are
subject to the general scaling law
$$\chi_{T/L}^{-1}(\tau,{\vec q},\omega) = q^{2 - \eta} \, {\hat
\chi}_{T/L}^{-1}({\vec q} \xi \, , \, \omega / \omega_c) \quad ,
\eqno(2.25)$$
where
$$\xi^{-1} = \mu \, m(1)^{2/(2 - \epsilon + \eta)} \propto \mid
\tau \mid^\nu \eqno(2.26a)$$
is the inverse static correlation length [remember that $m(1)
\propto {\bar \phi} \propto \mid \tau \mid^\beta$ and the scaling
relation $\nu = 2 \, \beta / (2 - \epsilon + \eta)$] with $\eta =
0$ to one-loop order, and
$$\omega_c = \lambda \mu^{2 - z} \, \xi^z \propto \mid \tau
\mid^{z \, \nu} \eqno(2.26b)$$
denotes the characteristic frequency scale. The asymptotic power
laws for the coexistence singularities are fixed by the anomalous
dimension (2.21b); remarkably one finds to leading order for
${\vec q} \rightarrow 0$ and $\omega \rightarrow 0$
$$\chi_L({\vec q},\omega) \propto {\bar \phi}^2 \, \big \vert
\chi_T({\vec q},\omega) \big \vert^{\epsilon / 2} \quad .
\eqno(2.27)$$
This means that the longitudinal correlation function displays an
anomalous line-shape, considerably stretched on the frequency
scale in comparison with a Lorentzian, and corresponding to an
algebraic decay of $G_L({\vec x},t)$ for long times. Of course,
Eq.(2.24) can be specialized to the static limit ($\omega = 0$)
or the case ${\vec q} \rightarrow 0$, $\omega / q^a = {\rm
const}$, for which the resulting coexistence anomalies are
summarized in Table I. For more information about the crossover
features compare Figs.7-9 and the discussion in Ref.[11].
\vfill \eject

\centerline{\bf III. CROSSOVER TO AN ASYMPTOTICALLY GAUSSIAN}
\centerline{\bf THEORY: THE CASE $n = 1$}
\bigskip
In the case of a single-component order parameter there will be
no massless modes in the low-temperature phase, of course, and
hence infrared singularities are confined to the vicinity of
$T_c$. However, one expects not only the critical region itself,
but also the crossover into the asymptotic "Gaussian" regime to
display a universal character. As an example, we therefore
extrapolate our generalized renormalization scheme to the
situation $T < T_c$ for $n = 1$, because the formalism allows the
separation of the universal behavior from any features depending
on the specific initial values of the flow parameters. This
approach to the crossover towards a Gaussian model will turn out
to be a necessary prerequisite for an appropriate discussion of
the anisotropic $n$-component relaxational models.

Hence we put $n = 1$ in the flow equations (2.19) using the
explicit one-loop results (2.17). Again, the $m$-dependent term
describes the freezing-out of the fluctuations on leaving the
critical region. For $m \rightarrow \infty$ only the mean-field
contributions to Wilson's $\zeta$- and $\beta$-functions remain,
leading to a power-law behavior of the flow parameters according
to their canonical dimensions
$$\eqalignno{m(\ell) &\propto \ell^{- 1} &(3.1a) \cr
u(\ell) &\propto \ell^{- \epsilon} \quad . &(3.1b) \cr}$$
Thus the anharmonicity $u(\ell)$ diverges (see also Eq.(2.21a)
for $n \rightarrow 1$); however, for large
$m(\ell)$ the relevant effective coupling entering the flow
functions vanishes according to
$$u_{eff}(\ell) = {u(\ell) \over [1 + m(\ell)^2]^{1 + \epsilon /
2}} \propto \ell^2  \quad . \eqno(3.2)$$
This eventually leads to mean-field exponents, which is precisely
what should be expected for a Gaussian model. On the other hand,
the loop-corrections to the scaling functions are of the order $u
\, m^{- \epsilon} \propto \ell^0$, i.e.: a non-universal
constant. Hence the asymptotic perturbation theory, although
non-trivial with respect to the anharmonicity, does not contain
any infrared singularities. Altogether, this seems to provide a
consistent description for the crossover to an asymptotically
uncritical regime.

These considerations can be confirmed by the investigation of an
approximate solution of Eq.(2.19b) simply using (3.1a) for the
complete crossover region. Instead of Eqs.(4.26a),(4.27a) of
Ref.[11] one finds for $\epsilon > 1$
$$u(\ell) = {u \, \ell^{- \epsilon} \over 1 + {3 \, u \over 2 \,
\epsilon} \left[ (m^2 + \ell^2)^{- \epsilon / 2} - (m^2 +
1)^{- \epsilon / 2} \right]} \quad , \eqno(3.3a)$$
while in the case $\epsilon = 0$ the result is identical to
Eq.(4.27b) of Ref.[11] for $n = 1$
$$u(\ell) = {u \over 1 - {3 \over 4} \, u \, \ln {m^2 + \ell^2
\over m^2 + 1}} \quad . \eqno(3.3b)$$
Hence at $d_c = 4$ even the logarithmic corrections in the
ordered phase disappear, and anomalous exponents are confined to
the immediate vicinity of the critical point. Note that according
to (3.1a) $m(\ell)$ diverges for $\epsilon = 2$, too, which
implies that the above mentioned inconsistency when assuming a
phase with spontaneously broken continuous symmetry for $n \geq
2$ at two dimensions, does not appear in the Ising case. Of
course, contrary to the systems where the Mermin-Wagner-Hohenberg
theorem [22,23] applies, the two-dimensional Ising model displays
a second-order phase transition at a finite temperature.

Fig.1a shows the numerical flow diagram for the effective
coupling $u_{eff}(\ell)$ vs. the quantity $m(\ell)^2 / [1 +
m(\ell)^2]$; compare with Fig.4 of Ref.[11]. Again, for $m(1) <
0.1$ one finds scaling behavior with respect to the variable $x$
of Eq.(2.23), which is not surprising as this is essentially a
property of the critical point. Hence for temperatures not too
far away from $T_c$ universal crossover features can be studied,
see Fig.1b (and compare Fig.5 of Ref.[11]), starting from the
Ising fixed point
$$\eqalignno{u_I^* &= {2 \, \epsilon \over 3} &(3.4a)\cr
\zeta_{m \, I}^* &= - 2 + \epsilon &(3.4b)\cr}$$
and terminating at the asymptotically stable Gaussian fixed point
with
$$\eqalignno{u_G^* = \infty \quad , &\qquad u_{eff}^* = 0
&(3.5a)\cr
\zeta_{m \, G}^* &= - 2 \quad . &(3.5b)\cr}$$

We illustrate the ensuing crossover within the ${\vec q}$- and
$\omega$-dependence of the correlation functions for the case of
the static susceptibility. Using the numerical solutions for
$u(\ell)$ and $m(\ell)$, and inserting $n = 1$ into Eq.(5.11b) of
Ref.[11], one arrives at the one-loop cumulant (C). The
corresponding vertex function (V) and the zero-loop expression
follow by expanding with respect to $u$ or putting $u = 0$,
respectively. For comparison of the different results the
effective exponent
$$2 - \eta_{eff} = {\partial \ln \chi^{-1}({\vec q},0) \over
\partial \ln q} \eqno(3.6)$$
is plotted in Fig.2. In all three cases shown, one finds similar
features because grossly the flow of the parameter $m(\ell)$
describes most of the crossover behavior. For large values of $q$
one has $\eta_{eff} = 0$ (to one-loop order), before eventually
the static susceptibility becomes finite for ${\vec q}
\rightarrow 0$. This is to be contrasted with the divergences
induced by the transverse massless modes for the models with $n
\geq 2$ (see Fig.7 with $y = 0$). We finally remark that our
interpretation for the minimum of the effective exponent $2 -
\eta_{eff}^L$ apparent for the time-dependent Heisenberg models
at intermediate wavenumbers [11], namely that it is due to the
freezing-out of the longitudinal fluctuations before the
Goldstone modes become the predominant excitations, is
quantitatively supported by Fig.2.
\bigskip \bigskip

\centerline{\bf IV. THE INFLUENCE OF WEAK CUBIC ANISOTROPIES ON
THE}
\centerline{\bf CRITICAL DYNAMICS OF MODELS A AND B BELOW $T_c$}
\bigskip

\centerline{\bf A. Model and renormalization}
\medskip
We now start our investigations of the influence of
characteristic anisotropies with the discussion of the crossover
properties when cubic anharmonicities have to be taken into
account. Thus we add the term
$$\Delta H_{cub}[\{ \phi_0^\alpha \}] = \int d^dx \, {v_0 \over
4!} \sum_{\alpha = 1}^n {\phi_0^\alpha({\vec x})}^4 \eqno(4.1)$$
to the Hamiltonian (2.1a). Its static properties above and below
$T_c$ were examined by Wallace [18] and Aharony [24] on the basis
of a $1/n$-expansion, and by Ketley and Wallace [25] using the
$\epsilon$-expansion. The phase diagram of the cubic model was
described in more detail by Br\'ezin, Le Guillou and Zinn-Justin
[26], Nattermann and Trimper [27], Lyuksyutov and Pokrovsky [28],
Rudnick [29], Iacobson and Amit [30]. For a summary of their main
results we refer to Chap.4 of Ref.[20]. We shall restrict
ourselves to the case $v_0 < 0$, for then the spontaneous order
parameter stays along one of the principal axes of the
$n$-dimensional hypercube. For stability reasons we furthermore
have to require
$$b_0 = u_0 + v_0 > 0 \quad ; \eqno(4.2a)$$
for $b_0 = 0$ the transition becomes first-order [27-30]. Finally
we are only interested in a small deviation from the
$O(n)$-symmetric case and hence tiny anisotropy parameters
$$0 \leq y_0 = {- v_0 \over b_0} \ll 1 \quad , \eqno(4.2b)$$
such that the continuous character of the transition is
guaranteed [28].

Eq.(2.3) again constitutes the Langevin dynamics of the system,
supplemented by the moments (2.4) for the stochastic forces. For
the spontaneous order parameter of Eq.(2.9a), however, we write
$${\bar \phi}_0 = \sqrt{3 \over b_0} \, m_0 \eqno(4.3)$$
instead of (2.10), and the fluctuation-induced $T_c$-shift as
determined from Eq.(2.11) for $m_0 = 0$ acquires an additional
contribution $\propto v_0$
$$r_{0c} = \left( {n + 2 \over 6 \, \epsilon} \, u_0 \, A_d + {1
\over 2 \, \epsilon} \, v_0 \, A_d \right)^{2 / \epsilon}
\eqno(4.4)$$
with respect to the isotropic case.

Hence in the ordered phase one finds for the cubic Janssen-De
Dominicis functional $J = J_0 + J_{int} + J_{CT}$, where the
harmonic part explicitly reads
$$\eqalignno{J_0[\{ {\tilde \pi}_0^\alpha \} , {\tilde \sigma}_0
, \{ \pi_0^\alpha \} , \sigma_0] = \int_q \int_\omega &\biggl[
\sum_\alpha \lambda_0 \, q^a \, {\tilde \pi}_0^\alpha({\vec
q},\omega) \, {\tilde \pi}_0^\alpha(- {\vec q},-\omega) +
\lambda_0 \, q^a \, {\tilde \sigma}_0({\vec q},\omega) \, {\tilde
\sigma}_0(- {\vec q},-\omega) \cr
&- \sum_\alpha {\tilde \pi}_0^\alpha({\vec q},\omega) \, \Bigl[ i
\omega + \lambda_0 \, q^a \, ({\overline m}_0^2 + q^2) \Bigr] \,
\pi_0^\alpha(- {\vec q},- \omega) \cr
&- {\tilde \sigma}_0({\vec q},\omega) \, \Bigl[ i \omega +
\lambda_0 \, q^a \, (m_0^2 + q^2) \Bigr] \, \sigma_0(-
{\vec q},- \omega) \biggr] \; , &(4.5a)\cr}$$
while the interactions are given by
$$\eqalignno{J_{int}&[\{ {\tilde \pi}_0^\alpha \} , {\tilde
\sigma}_0 , \{ \pi_0^\alpha \} , \sigma_0] = - {1 \over 6} \,
\lambda_0 \, \int_{q_1 q_2 q_3 q_4} \int_{\omega_1 \omega_2
\omega_3 \omega_4} q_1^a \, \delta \! \left( \sum_i {\vec q}_i
\right) \, \delta \! \left( \sum_i \omega_i \right) \times \cr
&\qquad \times \biggl[ \sum_{\alpha \beta} u_0 \, {\tilde
\pi}_0^\alpha({\vec q}_1,\omega_1) \, \pi_0^\alpha({\vec
q}_2,\omega_2) \, \pi_0^\beta({\vec q}_3,\omega_3) \,
\pi_0^\beta({\vec q}_4,\omega_4) \cr
&\qquad \qquad \qquad + \sum_\alpha v_0 \, {\tilde
\pi}_0^\alpha({\vec q}_1,\omega_1) \, \pi_0^\alpha({\vec
q}_2,\omega_2) \, \pi_0^\alpha({\vec q}_3,\omega_3) \,
\pi_0^\alpha({\vec q}_4,\omega_4) \cr
&\qquad \qquad \qquad + \sum_\alpha u_0 \, {\tilde
\pi}_0^\alpha({\vec q}_1,\omega_1) \, \pi_0^\alpha({\vec
q}_2,\omega_2) \, \sigma_0({\vec q}_3,\omega_3) \, \sigma_0({\vec
q}_4,\omega_4) \cr
&\qquad \qquad \qquad + \sum_\alpha u_0 \, {\tilde
\sigma}_0({\vec q}_1,\omega_1) \, \pi_0^\alpha({\vec
q}_2,\omega_2) \, \pi_0^\alpha({\vec q}_3,\omega_3) \,
\sigma_0({\vec q}_4,\omega_4) \cr
&\qquad \qquad \qquad \qquad + b_0 \, {\tilde \sigma}_0({\vec
q}_1,\omega_1) \, \sigma_0({\vec q}_2,\omega_2) \, \sigma_0({\vec
q}_3,\omega_3) \, \sigma_0({\vec q}_4,\omega_4) \biggr]
&(4.5b)\cr
&- {1 \over 6} \, \lambda_0 \, \sqrt{3 \over b_0} \, m_0
\int_{q_1 q_2 q_3} \int_{\omega_1 \omega_2 \omega_3} q_1^a \,
\delta \! \left( \sum_i {\vec q}_i \right) \, \delta \! \left(
\sum_i \omega_i \right) \times \cr
&\times \biggl[ \sum_\alpha 2 \, u_0 \, {\tilde
\pi}_0^\alpha({\vec q}_1,\omega_1) \, \pi_0^\alpha({\vec
q}_2,\omega_2) \, \sigma_0({\vec q}_3,\omega_3) \cr
&\qquad + \sum_\alpha u_0 \, {\tilde \sigma}_0({\vec
q}_1,\omega_1) \, \pi_0^\alpha({\vec q}_2,\omega_2) \,
\pi_0^\alpha({\vec q}_3,\omega_3) + 3 \, b_0 \, {\tilde
\sigma}_0({\vec q}_1,\omega_1) \, \sigma_0({\vec q}_2,\omega_2)
\, \sigma_0({\vec q}_3,\omega_3) \biggr] \; , \cr}$$
and the counter-term stemming from the equation of state (2.11)
finally is
$$\eqalignno{J_{CT}[\{ {\tilde \pi}_0^\alpha \} , {\tilde
\sigma}_0 , \{ \pi_0^\alpha \} , \sigma_0] &= - \lambda_0 \, A \!
\int_q \int_\omega q^a \Bigl[ \sum_\alpha {\tilde
\pi}_0^\alpha({\vec q},\omega) \pi_0^\alpha(-{\vec q},-\omega) +
{\tilde \sigma}_0({\vec q},\omega) \sigma_0(-{\vec q},-\omega)
\Bigr] \cr
&\qquad - \lambda_0 \, \sqrt{3 \over b_0} \, m_0 \, A \, {\tilde
\sigma}_0(0,0) \, \delta_{a,0} \quad . &(4.5c)\cr}$$
The most important difference to the isotropic model, which of
course emerges as the special case with $v_0 = 0$ and $b_0 =
u_0$, is the appearance of the transverse mass (remember $v_0 <
0$)
$${\overline m}_0^2 = {- v_0 \over 2 b_0} \, m_0^2 \eqno(4.6)$$
in the harmonic functional (4.5a). The basic elements of the
perturbation theory to be constructed from Eq.(4.5) and their
graphic representations are depicted in Fig.3.

The ultraviolet divergences of the dynamic correlation functions
are multiplicatively renormalized by introducing the $Z$-factors
$$\eqalignno{m^2 &= Z_m^{-1} \, m_0^2 \, \mu^{-2} &(4.7a)\cr
{\overline m}^2 &= Z_{\overline m}^{-1} \, {\overline m}_0^2 \,
\mu^{-2} &(4.7b)\cr
b &= Z_b^{-1} \, b_0 \, A_d \, \mu^{- \epsilon} &(4.7c) \cr
v &= Z_v^{-1} \, v_0 \, A_d \, \mu^{- \epsilon} &(4.7d) \cr}$$
in analogy to (2.15). In the framework of a one-loop theory we
may omit the field renormalizations and hence the $Z$-factor for
the Onsager coefficient $\lambda$. The renormalization constants
introduced in (4.7) are not independent; e.g. from the definition
(4.3) one easily derives
$$Z_b = Z_m \, Z_\sigma \quad , \eqno(4.8a)$$
with $Z_\sigma = 1$ to one-loop order. Similarly, (4.6) leads to
$$Z_b \, Z_{\overline m} = Z_v \, Z_m \quad . \eqno(4.8b)$$

The renormalization constants (4.7a,b) are conveniently
determined by rendering the quantities $\left( \partial_{q^2}
\right)^{a/2} \, \Gamma_{{\tilde \sigma} \sigma}({\vec
q},\omega)$ and $\left( \partial_{q^2} \right)^{a/2} \,
\Gamma_{{\tilde \pi} \pi}({\vec q},\omega)$, respectively, finite
at the normalization point $q = \mu$, $\omega = 0$. Using the
explicit expressions for the two-point vertex functions listed in
the Appendix, the one-loop results are
$$Z_m = Z_b = 1 + {n - 1 \over 6 \, \epsilon} \, {(b_0 - v_0)^2
\over b_0} \, {A_d \, \mu^{- \epsilon} \over \left( 1 +
{\overline m}_0^2 / \mu^2 \right)^{\epsilon / 2}} + {3 \over 2 \,
\epsilon} \, {b_0 \, A_d \, \mu^{- \epsilon} \over \left( 1 +
m_0^2 / \mu^2 \right)^{\epsilon / 2}} \eqno(4.9a)$$
$$Z_{\overline m} = Z_v = 1 + {1 \over 2 \, \epsilon} \, {4 \,
b_0 - v_0 \over 2 \, b_0 + v_0} \, {v_0 \, A_d \, \mu^{-
\epsilon} \over \left( 1 + {\overline m}_0^2 / \mu^2
\right)^{\epsilon / 2}} + {1 \over \epsilon} \, {4 \, b_0 - v_0
\over 2 \, b_0 + v_0} \, {b_0 \, A_d \, \mu^{- \epsilon} \over
\left( 1 + m_0^2 / \mu^2 \right)^{\epsilon / 2}} \quad .
\eqno(4.9b)$$
Obviously, for $v_0 = 0$ (4.9a) reduces to Eq.(4.11c) of Ref.[11]
for the isotropic model.

The $\zeta$- and $\beta$-functions following from (4.9) read
$$\zeta_m = \mu \, {\partial \over \partial \mu} \Big \vert_0 \,
\ln {m^2 \over m_0^2} = - 2 + {n - 1 \over 6} \, {(b - v)^2 / b
\over (1 - {v \over 2 \, b} \, m^2)^{1 + \epsilon / 2}} + {3
\over 2} \, {b \over (1 + m^2)^{1 + \epsilon / 2}} \eqno(4.10a)$$
$$\beta_b = \mu \, {\partial \over \partial \mu} \Big \vert_0 \,
b = b \left[ - \epsilon + {n - 1 \over 6} \, {(b - v)^2 / b
\over (1 - {v \over 2 \, b} \, m^2)^{1 + \epsilon / 2}} + {3
\over 2} \, {b \over (1 + m^2)^{1 + \epsilon / 2}} \right]
\eqno(4.10b)$$
$$\beta_v = \mu \, {\partial \over \partial \mu} \Big \vert_0 \,
v = v \left[ - \epsilon + {4 \, b - v \over 2 \, b + v} \left( {v
/ 2 \over (1 - {v \over 2 \, b} \, m^2)^{1 + \epsilon / 2}} + {b
\over (1 + m^2)^{1 + \epsilon / 2}} \right) \right] \; ,
\eqno(4.10c)$$
and the flow-dependent couplings are given by the following set
of ordinary diffential equations
$$\eqalignno{\ell \, {d m(\ell) \over d \ell} &= {1 \over 2} \,
m(\ell) \, \zeta_m(\ell) &(4.11a) \cr
\ell \, {d b(\ell) \over d \ell} &= \beta_b(\ell) &(4.11b) \cr
\ell \, {d v(\ell) \over d \ell} &= \beta_v(\ell) \quad ,
&(4.11c) \cr}$$
with the usual initial conditions $m(1) = m$, $b(1) = b$ and
$v(1) = v$. The renormalization group equations for the two-point
cumulants are finally solved by [compare Eq.(2.18b)]
$$G^c_{{\tilde N} N}(\mu , \lambda , m , b , v ,\{ {\vec q} \},\{
\omega \}) = G^c_{{\tilde N} N}(\mu \ell , \lambda , m(\ell) ,
b(\ell) , v(\ell) , \{ {\vec q} / \mu \ell \},\{ \omega / (\mu
\ell)^2 \}) \; . \eqno(4.12)$$
\bigskip

\centerline{\bf B. Discussion of the flow equations and fixed
points}
\medskip
The common zeros of the $\beta$-functions (4.10b,c) yield the
fixed points governing the universal critical behavior
$$\beta_b(b^*,v^*,m) = 0 = \beta_v(b^*,v^*,m) \quad .
\eqno(4.13)$$
Their regions of stability can be inferred by investigation of
the matrix
$$\Omega = \pmatrix{\partial \beta_b / \partial b & \partial
\beta_b / \partial v \cr
\partial \beta_v / \partial b & \partial \beta_v / \partial v
\cr} \bigg \vert_{b = b^* , \, v = v^*} \eqno(4.14)$$
(with fixed $m$): if both eigenvalues of $\Omega$ are positive,
the fixed point is (infrared-)stable, one positive and one
negative eigenvalue correspond to a saddlepoint, and two negative
eigenvalues lead to instability.

We begin with the case $m = 0$, i.e.: the properties of the
critical point $T = T_c$. Besides the Gaussian fixed point $b_G^*
= v_G^* = 0$, which is unstable below $d_c = 4$ because both
eigenvalues of $\Omega_G$ are equal to $- \epsilon$, one finds
three non-trivial solutions of Eq.(4.13), namely (i) the
isotropic Heisenberg fixed point (see Eq.(2.20a) of Section II.B)
$$b^*_H = {6 \, \epsilon \over n + 8} \quad , \qquad v^*_H = 0
\eqno(4.15a)$$
$$\Omega_H = \pmatrix{1 & {2 \, (n - 1) \over n + 8} \cr
0 & {4 - n \over n + 8} \cr} \epsilon \quad , \eqno(4.15b)$$
(ii) the Ising fixed point [compare Eq.(3.4a)]
$$b^*_I = v^*_I = {2 \, \epsilon \over 3} \eqno(4.16a)$$
$$\Omega_I = \pmatrix{1 & 0 \cr - \, {4 \over 3} & - \, {1 \over
3} \cr} \epsilon \eqno(4.16b)$$
with saddlepoint-type behavior, and finally (iii) the cubic fixed
point
$$b^*_A = {2 \, \epsilon \over 3} \, {n - 1 \over n} \quad ,
\qquad v^*_A = {2 \, \epsilon \over 3} \, {n - 4 \over n}
\eqno(4.17a)$$
$$\Omega_A = \pmatrix{5 \, n - 8 & 2 \, (n - 1) \cr
4 \, (4 - n) & 4 - n \cr} {\epsilon \over 3 \, n} \quad ,
\eqno(4.17b)$$
where $\Omega_A$ has the eigenvalues $\epsilon$ and $(n - 4) \,
\epsilon / 3 \, n$ [20 (and references therein)]. Therefore
the situations $n \geq n_c$ and $n < n_c$ have to be
distinguished, where to one-loop order $n_c = 4$ (in more
elaborate calculations one finds that $n_c$ is between $3$ and
$4$) [25-27]. For $n \geq n_c$ the cubic fixed point (4.17a)
is stable, and hence the anisotropy (4.1) constitutes a relevant
perturbation of the $O(n)$-symmetric Hamiltonian (2.1); to the
contrary, in the case $n < n_c$ the Heisenberg fixed point (4.15)
becomes asymptotically stable, which means that the fluctuations
dynamically restore the rotational symmetry. (An interpretation
for this remarkable difference might be that in high-dimensional
spaces most of the "volume" is concentrated in the vicinity of
the surface, while for low dimensionality $n < n_c$ the detailed
shape of the surface in configuration space is less relevant and
the "edges" are effectively smoothed by the fluctuations.) For $n
= 4$ Eqs.(4.15) and (4.17) are identical, and of course for $n =
1$ only the Ising fixed point (4.16) remains and becomes stable,
compare Section III. In the limit $n \rightarrow \infty$ (4.16a)
and (4.17a) coincide.

{}From (4.15)-(4.17) we can deduce the qualitative features of the
flow diagrams, which are displayed in Fig.4 for the typical cases
$n = 2$ and $n = 8$, respectively. In Fig.4a the attractivity of
the Heisenberg fixed point "H" (4.15) is evident, whereas in
Fig.4b the anisotropic cubic fixed point "A" (4.17) dominates the
flow of the couplings. Those trajectories which are not located
in the regime of the respective stable fixed point eventually
cross the stability boundary $b = 0$ and the corresponding
physical system will then display a first-order phase transition
[28-30,20].

We can now turn to the phase with spontaneously broken (discrete
cubic) symmetry, i.e.: the case $m \rightarrow \infty$. Apart
from the (for $\epsilon > 0$) unstable Gaussian fixed point
$b_G^* = v_G^* = 0$ there is now the isotropic coexistence fixed
point [see (2.21a)]
$$b^*_C = {6 \, \epsilon \over n - 1} \quad , \qquad v^*_C = 0
\eqno(4.18a)$$
$$\Omega_C = \pmatrix{1 & 2 \cr 0 & - 1 \cr} \epsilon \quad ,
\eqno(4.18b)$$
which, however, in contrast to the isotropic model behaves like a
saddlepoint in parameter space. As long as $v \not= 0$, we have
asymptotically ${\overline m}(\ell) \rightarrow \infty$. Hence
from (4.10) and (4.11) we conclude
$$\eqalignno{m(\ell) &\propto \ell^{- 1} &(4.19a) \cr
b(\ell) &\propto \ell^{- \epsilon} &(4.19b) \cr
v(\ell) &\propto \ell^{- \epsilon} \quad , &(4.19c) \cr}$$
i.e.: the mass parameter and anharmonic couplings diverge in a
manner completely analogous to what we have previously discussed
for the case $n = 1$. Of course, a crossover to an asymptotically
Gaussian theory should in fact be expected because the
fluctuations are bound to cease eventually. This can be
confirmed by studying the effective couplings
$$\eqalignno{b_{eff}(\ell) &= {b(\ell) \over [1 + {\overline
m}(\ell)^2]^{1 + \epsilon / 2}} \propto \ell^2 &(4.20a)\cr
v_{eff}(\ell) &= {v(\ell) \over [1 + {\overline m}(\ell)^2]^{1 +
\epsilon / 2}} \propto \ell^2 \quad , &(4.20b) \cr}$$
compare Section III and Fig.5. In particular, we want to point
out that in the presence of cubic anharmonic terms the
Mermin-Wagner-Hohenberg theorem [22,23] does not apply and hence
long-range order is possible in two dimensions, which is
reflected in the $d$-independent divergence of the mass parameter
according to (4.19a) instead of the anomalous dimension (2.21b)
of the isotropic model.

The fact that (to one-loop order) for $n < 4$ the
$O(n)$-symmetry is dynamically restored at the critical point
turns out to be most important for the flow of the effective
coupling $b_{eff}(\ell)$ and hence the static and dynamic
susceptibilities, at least for low initial values of the
anisotropy parameter $y$ of Eq.(4.2b). Ideally the system would
acquire $v_H^* = 0$ at $T_c$ leading to pure coexistence behavior
as described in Section II. However, it is unstable towards any
small disturbances in the $v$-direction followed by a crossover
into the asymptotically Gaussian model. We may study this
situation quantitatively by changing the initial value $y(1)$
in the numerical solution of the flow equations (4.10), (4.11).
In Fig.5 the flow of the effective coupling (4.20a) vs. the
scaling variable (2.23) is depicted for several values of $y(1)$;
for comparison the crossover from the Heisenberg to Lawrie's
coexistence fixed point in the $O(n)$-symmetric model is plotted,
too. For very low anisotropy $b_{eff}(\ell)$ rises towards the
coexistence fixed point (4.18), but finally approaches the
Gaussian limit [see Eq.(3.5)]. The smaller $y(1)$, the higher is
the maximum of $b_{eff}(\ell)$, and the later the crossover to
$b_{eff}^* = 0$ takes place. In fact the approach to
$b_{eff}(\ell) \rightarrow 0$ could be universally rescaled by
using the "Gaussian scaling variable" $\ell / {\overline m}(1)$
instead of (2.23).

The large attracting region of the isotropic coexistence fixed
point (4.18) becomes obvious from the qualitative flow diagram in
Fig.6a (with $n = 2$; compare Fig.4a). For $n > 4$, on the other
hand, for the allowed values $v < 0$ the Gaussian limit is
immediately assumed, without any relics of the coexistence
region, see Fig.6b. Even for very low values of $y(1)$ no
quasi-Goldstone modes are of importance. We finally remark that
the above considerations, especially for the critical region $m =
0$, are inferred from the one-loop approximation only. Thus by
taking into account higher orders of the perturbation expansion
modifications might be necessary. In particular, the number pf
components $n_c$ below which the Heisenberg fixed point is the
attractive critical fixed point is shifted. It should be clear,
though, how the properties of the critical point imply the
asymptotic features for $T < T_c$ and ${\vec q} \rightarrow 0$,
$\omega \rightarrow 0$.
\bigskip

\centerline{\bf C. Dynamic response and correlation functions}
\medskip
With the help of Eq.(4.12) we can now evaluate the dynamic
two-point cumulants to one-loop order from the diagrams listed in
the Appendix. We furthermore apply the general matching condition
(2.24) designed to connect the low-wavenumber and low-frequency
behavior with the asymptotic limit $\ell \rightarrow 0$ where the
stable fixed points are approached. In the presence of cubic
anisotropies the perturbation series cannot be truncated at the
one-loop level, however, as opposed to the isotropic case. For
this reason, and to avoid somewhat even lengthier expressions, we
quote the results for the one-loop vertex functions which may be
easily transformed into the cumulants for comparison with the
$O(n)$-symmetric models [11]. Specializing (2.24) to $\omega =
0$, using Eq.(4.9) and the abbreviation
$$S(\ell) = \sqrt{1 + {2 \, b(\ell) - v(\ell) \over b(\ell)} \,
m(\ell)^2 + {[2 \, b(\ell) + v(\ell)]^2 \over 4 \, b(\ell)^2} \,
m(\ell)^4} \quad , \eqno(4.21)$$
the inverse static susceptibilities at $d = 3$ ($\epsilon = 1$)
read (see Eq.(5.5) of Ref.[11])
$$\eqalignno{&\chi_T^{-1}({\vec q},0) = \mu^2 \, \ell^2 \Biggl[ 1
+ {- v(\ell) \over 2 \, b(\ell)} \, m(\ell)^2 \, Z_{\overline
m}(\ell) + {2 \, b(\ell) + v(\ell) \over 6} \! \left( 1 - \sqrt{-
v(\ell) \over 2 \, b(\ell)} \right) \! m(\ell) &(4.22a) \cr
&- {[b(\ell) - v(\ell)]^2 \over 6 \, b(\ell)} m(\ell)^2 \!
\left( \! \arcsin {1 - {2 \, b(\ell) + v(\ell) \over 2 \,
b(\ell)} \, m(\ell)^2 \over S(\ell)} + \arcsin {1 + {2 \, b(\ell)
+ v(\ell) \over 2 \, b(\ell)} \, m(\ell)^2 \over S(\ell)} \right)
\! \Biggr] \! \Bigg \vert_{\ell = q / \mu} \cr}$$
and
$$\eqalignno{\chi_L^{-1}({\vec q},0) = \mu^2 \, \ell^2 \, \Biggl[
1 &+ m(\ell)^2 \, Z_m(\ell) - {3 \over 2} \, b(\ell) \, m(\ell)^2
\, \arcsin {1 \over \sqrt{1 + 4 \, m(\ell)^2}} &(4.22b)\cr
&- {n - 1 \over 6} \, {[b(\ell) - v(\ell)]^2 \over b(\ell)} \,
m(\ell)^2 \, \arcsin {1 \over \sqrt{1 + {- 2 \, v(\ell) \over
b(\ell)} \, m(\ell)^2}} \Biggr] \Bigg \vert_{\ell = q / \mu}
\quad . \cr}$$

Again, their features are most conveniently explored by
introducing effective exponents
$$2 - \eta_{eff}^{T/L} = {\partial \ln \chi_{T/L}^{-1}({\vec
q},0) \over \partial \ln q} \quad , \eqno(4.23)$$
which are displayed in Fig.7 for $n = 2$ and the same values for
the anisotropy parameter $y(1)$ that have been used in Fig.5,
including $y(1) = 0$. (In order to allow for a most direct
comparison with the isotropic model, here the two-point cumulants
are plotted.) The transverse exponent (Fig.7a) shows how the
quasi-Goldstone modes for $y(1) \not= 0$ finally acquire a mass
and hence the transverse susceptibility becomes finite for ${\vec
q} \rightarrow 0$. If we had plotted $2 - \eta_{eff}^T$ against
$q / \mu \, {\overline m}(1)$ instead of the critical scaling
variable following from (2.23), we would have found a universal
crossover signature very similar to the one depicted in Fig.2 for
the one-component case. For very low anisotropies, on the other
hand, coexistence-like anomalies are clearly noticeable in the
wavevector-dependence of the longitudinal response function
within an intermediate $q$-region the size of which depends on
$y(1)$, if $n < n_c$. For $y(1) < 10^{-4}$, the pronounced
minimum of the effective exponent $2 - \eta_{eff}^L$ is then
still apparent, while for larger anisotropy the flow becomes much
more similar to Fig.2 for the direct crossover to the Gaussian
model. Now turning to $n \geq n_c$, according to the preceding
Subsection, no remnants of the Goldstone modes are to be
expected.

As in the isotropic case, the dynamic response functions may be
explicitly written down for arbitrary dimensions $2 < d \leq 4$
in the limit ${\vec q} \rightarrow 0$. For model A [1] with
non-conserved order parameter one finds (compare Eq.(5.11) of
Ref.[11])
$$\eqalignno{&\chi_T^{-1}(0,\omega) = \mu^2 \, \ell^2 \, \Biggl(
- i + {- v(\ell) \over 2 \, b(\ell)} \, m(\ell)^2 \, Z_{\overline
m}(\ell) &(4.24a)\cr
&\qquad + {2 \, b(\ell) + v(\ell) \over 6 \, \epsilon}
\left[ 1 - \left( {- v(\ell) \over 2 \, b(\ell)} \right)^{1 -
\epsilon / 2} \right] \, m(\ell)^{2 - \epsilon} - {[b(\ell) -
v(\ell)]^2 \over 3 \, \epsilon \, b(\ell)} \times \cr
&\times \Biggl[ {m(\ell)^2 \over i - {2 \, b(\ell) + v(\ell)
\over 2 \, b(\ell)} \, m(\ell)^2} \left[ \left( {- v(\ell) \over
2 \, b(\ell)} \, m(\ell)^2 \right)^{1 - \epsilon / 2} - \left( {-
i \over 2} + {2 \, b(\ell) - v(\ell) \over 4 \, b(\ell)} \,
m(\ell)^2 \right)^{1 - \epsilon / 2} \right] \cr
&\, + {m(\ell)^2 \over i + {2 \, b(\ell) + v(\ell) \over 2 \,
b(\ell)} \, m(\ell)^2} \left[ m(\ell)^{2 - \epsilon} - \left( {-
i \over 2} + {2 \, b(\ell) - v(\ell) \over 4 \, b(\ell)} \,
m(\ell)^2 \right)^{1 - \epsilon / 2} \right] \Biggr] \Biggr)
\Bigg \vert_{\ell = \sqrt{\omega / \lambda \mu^2}} \cr}$$
$$\eqalignno{\chi_L^{-1}&(0,\omega) = \mu^2 \, \ell^2 \Biggl( - i
+ m(\ell) \, Z_m(\ell) + {3 \, i \over \epsilon} \, b(\ell) \,
m(\ell)^2 \left[ m(\ell)^{2 - \epsilon} - \left( {- i \over 2} +
m(\ell)^2 \right)^{1 - \epsilon / 2} \right] \cr
&\qquad + {(n - 1) \, i \over 3 \, \epsilon} \, {[b(\ell) -
v(\ell)]^2 \over b(\ell)} \, m(\ell)^2 \times &(4.24b)\cr
&\qquad \times \left[ \left( {- v(\ell) \over 2 \, b(\ell)} \,
m(\ell)^2 \right)^{1 - \epsilon / 2} - \left( {- i \over 2} + {-
v(\ell) \over 2 \, b(\ell)} \, m(\ell)^2 \right)^{1 - \epsilon /
2} \right] \Biggr) \Bigg \vert_{\ell = \sqrt{\omega / \lambda
\mu^2}} \quad . \cr}$$
The logarithmic derivatives of the inverse real part of the
longitudinal susceptibility and of the inverse longitudinal
correlation function obtained with the help of (2.8) are depicted
for $n = 2$ and $\epsilon = 1$ in Figs.8a and 8b, respectively.
For the relevant situation $n < 4$, their general features are
similar to that of the static effective exponent $2 -
\eta_{eff}^L$; however, the crossover to the asymptotic Gaussian
behavior takes place for values of the dynamic scaling variable
about an order of magnitude lower than in the static case, and
the influence of the cubic anisotropies is considerably reduced.

If the order parameter field is conserved (model B according to
Hohenberg and Halperin [1]), due to the diffusion pole the
ratio $\omega / q^2$ has to be kept fixed when the limit ${\vec
q} \rightarrow 0$ is performed. Then merely the zero-loop
contributions (to be supplied with the full one-loop flow
equations) persist [11]:
$$\eqalignno{\chi_T^{-1}(0,\omega / q^2) &= \mu^2 \, \ell^2 \,
\left[ - i + {- v(\ell) \over 2 \, b(\ell)} \, m(\ell)^2 \right]
\Big \vert_{\ell = \sqrt{\omega / \lambda \, q^2}} &(4.25a) \cr
\chi_L^{-1}(0, \omega / q^2) &= \mu \, \ell^2 \, \Bigl[ - i +
m(\ell)^2 \Bigr] \Big \vert_{\ell = \sqrt{\omega / \lambda \,
q^2}} \quad . &(4.25b) \cr}$$
The ensuing crossover to the Gaussian theory with pronounced
remnants of coexistence anomalies again is shown in Fig.9, for
the case $n = 2$ and $d = 3$.

For a comparison with the results (5.13) of Ref.[11] we finally
quote the complete ${\vec q}$- and $\omega$-dependent response
functions for model A at three dimensions
$$\eqalignno{&\chi_T^{-1}({\vec q},\omega) = q^2 - {i \, \omega
\over \lambda} + \mu^2 \, \ell^2 \, \Biggl[ {- v(\ell) \over 2 \,
b(\ell)} \, m(\ell)^2 \, Z_{\overline m}(\ell) + {2 \, b(\ell) +
v(\ell) \over 6} \, \left( 1 - \sqrt{- v(\ell) \over 2 \,
b(\ell)} \right) \, m(\ell) \cr
&\quad - {[b(\ell) - v(\ell)]^2 \over 6 \, b(\ell)} \, {m(\ell)^2
\over q / \mu \, \ell} \Biggl[ \arcsin \left({q^2 - i \, \omega /
\lambda \over \mu^2 \, \ell^2 \, S_+(\ell)} - {2 \, b(\ell) +
v(\ell) \over 2 \, b(\ell) \, S_+(\ell)} \, m(\ell)^2 \right)
&(4.26a)\cr
&+ \arcsin \left({i \omega / \lambda \over \mu^2 \ell^2
S_+(\ell)} + {2 b(\ell) + v(\ell) \over 2 b(\ell) S_+(\ell)}
m(\ell)^2 \right) + \arcsin \left({i \omega / \lambda \over \mu^2
\ell^2 S_-(\ell)} - {2 \, b(\ell) + v(\ell) \over 2 b(\ell)
S_-(\ell)} m(\ell)^2 \right) \cr
&\qquad \qquad \qquad \qquad \qquad \qquad + \arcsin \left({q^2 -
i \, \omega / \lambda \over \mu^2 \, \ell^2 \, S_-(\ell)} + {2 \,
b(\ell) + v(\ell) \over 2 \, b(\ell) \, S_-(\ell)} \, m(\ell)^2
\right) \Biggr] \Biggr] \cr}$$
and
$$\eqalignno{&\chi_L^{-1}({\vec q},\omega) = q^2 - {i \, \omega
\over \lambda} + \mu^2 \, \ell^2 \, \Biggl[ m(\ell)^2 \,
Z_m(\ell) \cr
&\quad - {n - 1 \over 6} {[b(\ell) - v(\ell)]^2 \over b(\ell)}
{m(\ell)^2 \over q / \mu \ell} \left( \arcsin {i \omega / \lambda
\over \mu^2 \ell^2 S_\perp(\ell)} + \arcsin {q^2 - i \omega /
\lambda \over \mu^2 \ell^2 S_\perp(\ell)} \right) &(4.26b)\cr
&\qquad \qquad - {3 \over 2} \, {b(\ell) \, m(\ell)^2 \over q /
\mu \, \ell} \left( \arcsin {i \, \omega / \lambda \over \mu^2 \,
\ell^2 \, S_\parallel(\ell)} + \arcsin {q^2 - i \, \omega /
\lambda \over \mu^2 \, \ell^2 \, S_\parallel(\ell)} \right)
\Biggr] \quad . \cr}$$
Here the abbreviations
$$S_\pm(\ell) = \sqrt{\left( {i \omega / \lambda \over \mu^2
\, \ell^2} \pm {2 b(\ell) + v(\ell) \over 2 b(\ell)} m(\ell)^2
\right)^2 \! + \left( {q^2 - 2 i \omega / \lambda \over \mu^2 \,
\ell^2} + {2 b(\ell) - v(\ell) \over b(\ell)} m(\ell)^2 \right)
{q^2 \over \mu^2 \, \ell^2}} \eqno(4.27a)$$
and
$$\eqalignno{S_\perp(\ell) &= \sqrt{\left( {i \, \omega / \lambda
\over \mu^2 \, \ell^2} \right)^2 + \left( {q^2 - 2 \, i \, \omega
/ \lambda \over \mu^2 \, \ell^2} + {- 2 \, v(\ell) \over b(\ell)}
\, m(\ell)^2 \right) \, {q^2 \over \mu^2 \, \ell^2}} &(4.27b)\cr
S_\parallel(\ell) &= \sqrt{\left( {i \, \omega / \lambda \over
\mu^2 \, \ell^2} \right)^2 + \left( {q^2 - 2 \, i \, \omega /
\lambda \over \mu^2 \, \ell^2} + 4 \, m(\ell)^2 \right) \, {q^2
\over \mu^2 \, \ell^2}} &(4.27c)\cr}$$
were used, and the matching condition
$$\ell^2 = \bigg \vert {q^2 \over \mu^2} - { i \, \omega \over
\lambda \, \mu^2} \bigg \vert \eqno(4.28)$$
has to be inserted. Investigation of the poles of (4.26) shows
that the damping of both the transverse and longitudinal modes
asymptotically is independent of $q$. Similarly, in the case of
diffusive relaxation (model B) the dispersion relation becomes $i
\, \omega \propto q^2$. The anomalous line shape (2.27) of the
longitudinal correlation function still appears in an
intermediate range, if $n \leq 4$ and $y(1) \ll 1$; finally it
develops into the regular Lorentz form.

The graphs discussed above refer to the situation where the
number of components is low, $n < n_c$. For larger $n$, the cubic
anisotropy immediately destroys the massless character of the
transverse excitations and hence no coexistence-type anomalies
persist. The theory is generally restricted to dimensionality $2
< d \leq 4$. At the upper critical dimension $d_c = 4$, instead
of the power laws logarithmic corrections come into play.
Although excluded from the scope of the formalism presented here,
in the limit $d \rightarrow 2$ reasonable statements may be
extracted from the general features of the flow equations, in the
sense that the results are compatible with exact theorems on the
existence of long-range order [22,23]. In Table II we summarize
the qualitative behavior under the influence of small cubic
anisotropies.
\bigskip \bigskip

\centerline{\bf V. THE INFLUENCE OF THE DIPOLAR INTERACTION ON
THE}
\centerline{\bf DYNAMICS OF MODEL A IN THE ORDERED PHASE}
\bigskip

\centerline{\bf A. Dipolar propagators in the ordered phase and}
\centerline{\bf crossover behavior at the critical point}
\medskip
The second very interesting generic anisotropy we want to study
is the dipol-dipol interaction. Although the phase transitions,
e.g. in magnetic systems, are driven by the short-range exchange
interactions, long-range dipolar forces become increasingly
important on approaching the critical point where the correlation
length diverges. Within the static renormalization group theory,
the influence of dipolar terms has been examined by Fisher and
Aharony [31], and was systematically treated together with other
kinds of anisotropies by Aharony [24]. The interplay between the
isotropic Heisenberg fixed point and an additional asympotically
stable dipolar fixed point produces remarkable crossover
phenomena, which have been intensively investigated in the
literature; we quote here the works of Bruce and Aharony [32],
Nattermann and Trimper [33], Bruce, Kosterlitz and Nelson [34],
and later of Santos [35], Kogon and Bruce [36], which are all
based on the renormalization group theory, and expecially the
recent papers by Frey and Schwabl [37] who applied the extended
renormalization scheme of Amit and Goldschmidt [14] to the
dipolar crossover problem.

Taking into account only those terms which are relevant with
respect to the renormalization group [31], the starting point of
our considerations is the following supplement to the static
Ginzburg-Landau functional (2.1)
$$\Delta H_{dip}[\{ \phi_0^\alpha \}] = \int_q \, {g_0 \over 2}
\, \sum_{\alpha , \beta = 1}^{{\rm min}(d,n)} {q_\alpha \,
q_\beta \over q^2} \, \phi_0^\alpha({\vec q}) \,
\phi_0^\beta(-{\vec q}) \quad . \eqno(5.1)$$
Hence the dipol-dipol interaction hence leads to a term
proportional to the coupling parameter $g_0$ which is anisotropic
in momentum space. We shall treat the general situation $n \not=
d$, although in part of our study we shall restrict ourselves to
the widely considered case of equal dimensions in momentum and
order parameter space. With the aid of the projectors [37]
$$\eqalignno{P^T_{\alpha \beta} &= \delta_{\alpha \beta} -
{q_\alpha \, q_\beta \over q^2} &(5.2a)\cr
P^L_{\alpha \beta} &= {q_\alpha \, q_\beta \over q^2} \quad ,
&(5.2b)\cr}$$
fulfilling the relations
$$\eqalignno{P^T_{\alpha \beta} + P^L_{\alpha \beta} &=
\delta_{\alpha \beta} &(5.3a)\cr
\sum_{\gamma = 1}^d P^{T/L}_{\alpha \gamma} \, P^{T/L}_{\gamma
\beta} &= P^{T/L}_{\alpha \beta} &(5.3b)\cr
\sum_{\gamma = 1}^d P^T_{\alpha \gamma} \, P^L_{\gamma \beta} &=
0 \quad , &(5.3c)\cr}$$
we may decompose the Hamiltonian into different contributions
corresponding to the transverse ($T$) and longitudinal parts
($L$) with respect to the wavevector ${\vec q}$
$$\eqalign{H_0[\{ \phi_0^\alpha \}] &= \int_q \Biggl( {1 \over 2}
\sum_{\alpha , \beta = 1}^{{\rm min}(d,n)} \Bigl[ (r_0 + q^2) \,
P^T_{\alpha \beta} + (r_0 + g_0 + q^2) \, P^L_{\alpha \beta}
\Bigr] \phi_0^\alpha({\vec q}) \, \phi_0^\beta(-{\vec q}) \cr
&\qquad \qquad + {1 \over 2} \, (r_0 + q^2) \sum_{\alpha =
{\rm min}(d,n) + 1}^n \phi_0^\alpha({\vec q}) \,
\phi_0^\alpha(-{\vec q}) \Biggr) \quad . \cr} \eqno(5.4)$$
{}From (5.4) it is apparent that the fluctuations in the
longitudinal sector acquire an additional mass term $\propto g_0$
due to the dipolar forces.

The static critical properties of any dipolar magnet are
described by the Hamiltonian (5.1) irrespective of the detailed
dynamics. For the latter we remark that in the presence of
dipolar forces the magnetization is no more a conserved quantity.
Hence the relaxation process would be that of model A [1],
Eq.(2.3) with $a = 0$. However, mode-coupling terms are most
important for the dynamic properties of ferromagnets and
antiferromagnets. Yet for the sake of simplicity, and because the
focus of our intention is rather the influence of the
anisotropies on the fluctuations, we shall drop the mode-coupling
vertex and restrict our investigations to the time-dependent
Ginzburg-Landau model A.

In order to avoid unnecessary complications, we start with the
important special situation where
$$n = d = 4 - \epsilon \quad , \eqno(5.5)$$
and shall come back to the general case later (Subsection C).
Below the transition temperature, we may still use the
parametrization (2.9), (2.10) for the spontaneous order
parameter. The Janssen-De Dominics functional (2.5b) can then be
split according to $J = J_0^T + J_0^L + J_{int} + J_{CT}$, with
the harmonic contributions
$$\eqalignno{J_0^T[&\{ {\tilde \pi}_0^\alpha \} , {\tilde
\sigma}_0 , \{ \pi_0^\alpha \} , \sigma_0] = &(5.6a)\cr
&= \int_q \int_\omega \biggl[ \sum_{\alpha \beta} \left(
\lambda_0 \, {\tilde \pi}_0^\alpha({\vec q},\omega) \, {\tilde
\pi}_0^\beta(- {\vec q},-\omega) - {\tilde \pi}_0^\alpha({\vec
q},\omega) \, \Bigl[ i \, \omega + \lambda_0 \, q^2 \Bigr] \,
\pi_0^\beta(- {\vec q},- \omega) \right) P^T_{\alpha \beta} \cr
&\quad + \left( \lambda_0 \, {\tilde \sigma}_0({\vec q},\omega)
\, {\tilde \sigma}_0(- {\vec q},-\omega) - {\tilde
\sigma}_0({\vec q},\omega) \, \Bigl[ i \, \omega + \lambda_0 \,
(m_0^2 + q^2) \Bigr] \, \sigma_0(-{\vec q},- \omega) \right)
P^T_{n n} \biggr] \cr}$$
and
$$\eqalignno{&J_0^L[\{ {\tilde \pi}_0^\alpha \} , {\tilde
\sigma}_0 , \{ \pi_0^\alpha \} , \sigma_0] = &(5.6b)\cr
&= \int_q \int_\omega \biggl[ \sum_{\alpha \beta} \left(
\lambda_0 \, {\tilde \pi}_0^\alpha({\vec q},\omega) \, {\tilde
\pi}_0^\beta(- {\vec q},-\omega) - {\tilde \pi}_0^\alpha({\vec
q},\omega) \, \Bigl[ i \, \omega + \lambda_0 \, (g_0 + q^2)
\Bigr] \, \pi_0^\beta(- {\vec q},- \omega) \right) P^L_{\alpha
\beta} \cr
&\qquad + \left( \lambda_0 \, {\tilde \sigma}_0({\vec q},\omega)
\, {\tilde \sigma}_0(- {\vec q},-\omega) - {\tilde
\sigma}_0({\vec q},\omega) \, \Bigl[ i \omega + \lambda_0 \,
(m_0^2 + g_0 + q^2) \Bigr] \, \sigma_0(-{\vec q},- \omega)
\right) P^L_{n n} \cr
&\qquad \quad - \sum_\alpha \lambda_0 \, g_0 \, \Bigl[ {\tilde
\pi}_0^\alpha({\vec q},\omega) \, \sigma_0(-{\vec q},-\omega)
+ {\tilde \sigma}_0({\vec q},\omega) \, \pi_0^\alpha(-{\vec q},-
\omega) \Bigr] P^L_{\alpha n} \biggr] \quad , \cr}$$
denoting the transverse and longitudinal parts with respect to
the wavevector ${\vec q}$. The interaction term $J_{int}$ is
identical to that of the isotropic model A and follows from
(4.5b) for $v_0 = 0$ and $b_0 = u_0$. Similarly, the counterterm
$J_{CT}$ looks like (4.5c), with $A$ replaced by the analogous
quantity ${\tilde A}$. In the transverse part (5.6a) of the
dynamic functional the $\pi$-modes are massless and in harmonic
approximation independent of the $\sigma$-fluctuation. The
longitudinal $\pi$-modes, on the other hand, have lost their
Goldstone character due to the dipolar interaction, and
furthermore couple to the $\sigma$-field ($\propto \lambda_0 \,
g_0$). We also remark that the operators $P_{\alpha \beta}^T$ and
$P_{\alpha \beta}^L$ are no proper projectors in the subspace of
the $n - 1$ fluctuations transverse with respect to the direction
of the order parameter.

The evaluation of the harmonic propagators is thus more difficult
than in the isotropic case. It is very convenient to collect the
response and correlation fields in a $2 \, n$-dimensional vector
according to ${\vec \psi}_0 = ({\tilde \pi}_0^1 , \ldots ,
{\tilde \pi}_0^{n-1} , {\tilde \sigma}_0 , \pi_0^1 , \ldots ,
\pi_0^{n-1} , \sigma_0)$. Then $J_0 = J_0^T + J_0^L$ can be
written as
$$J_0[{\vec \psi}_0] = - {1 \over 2} \int_q \! \int_\omega {\vec
\psi}_0({\vec q},\omega)^T \, {\hat A}({\vec q},\omega) \, {\vec
\psi}_0(- {\vec q},- \omega) \quad , \eqno(5.7a)$$
where the harmonic coupling matrix separates into
$n \times n$-submatrices
$${\hat A}({\vec q},\omega) = \pmatrix{{\hat D} & {\hat G} \cr
{\hat G}^* & 0 \cr} \quad . \eqno(5.7b)$$
Here ${\hat D}$ is diagonal
$${\hat D} = \pmatrix{- 2 \, \lambda_0 &  & \cr & \ldots & \cr &
& - 2 \, \lambda_0 \cr} \quad , \eqno(5.7c)$$
and the symmetric matrix ${\hat G}$ has the form
$${\hat G} = \pmatrix{B & v \cr v^T & b \cr} \quad ,
\eqno(5.7d)$$
with
$$\eqalignno{B_{\alpha \beta}({\vec q},\omega) &= (i \, \omega +
\lambda_0 \, q^2) \, \delta_{\alpha \beta} + \lambda_0 \, g_0 \,
{q_\alpha \, q_\beta \over q^2} &(5.8a)\cr
v_\alpha({\vec q}) &= \lambda_0 \, g_0 \, {q_\alpha \, q_n \over
q^2} &(5.8b)\cr
b({\vec q},\omega) &= i \, \omega + \lambda_0 \, (m_0^2 + q^2) +
\lambda_0 \, g_0 \, {q_n^2 \over q^2} \quad . &(5.8c)\cr}$$
The $(n-1) \times (n-1)$-matrix B is easily inverted:
$$B^{-1}_{\alpha \beta}({\vec q},\omega) = {1 \over i \, \omega +
\lambda_0 \, q^2} \left( \delta_{\alpha \beta} - {\lambda_0 \,
g_0 \over i \, \omega + \lambda_0 \, (g_0 + q^2) - \lambda_0 \,
g_0 \, q_n^2 / q^2} \, {q_\alpha \, q_\beta \over q^2}
\right) \quad . \eqno(5.9)$$
The inverse of ${\hat G}$ then follows as
$${\hat G}^{-1} = \pmatrix{C & u \cr u^T & c} \quad ,
\eqno(5.10a)$$
with
$$\eqalignno{c &= {1 \over b - v^T \, B^{-1} \, v} &(5.10b)\cr
u &= - c \, B^{-1} \, v &(5.10c)\cr
C &= B^{-1} \, (1 - v \, u^T) \quad . &(5.10d)\cr}$$
Finally one finds
$${\hat A}^{-1}({\vec q},\omega) = \pmatrix{0 & {\hat G}^{* \,
{-1}} \cr {\hat G}^{-1} & - {\hat G}^{-1} \, {\hat D} \, {\hat
G}^{* \, -1}} \quad , \eqno(5.11)$$
from which the propagators of the theory may be obtained (see
Ref.[11,12]). The zero-loop response propagators read
$$\eqalignno{G_{0 \, {\tilde \pi}^\alpha \pi^\beta}({\vec
q},\omega) &= {1 \over - i \, \omega + \lambda_0 \, q^2} \,
\left[ \delta_{\alpha \beta} - {q_\alpha \, q_\beta \over q^2} \,
s({\vec q},\omega) \right] &(5.12a)\cr
G_{0 \, {\tilde \pi}^\alpha \sigma}({\vec q},\omega) &= {- 1
\over - i \, \omega + \lambda_0 \, (m_0^2 + q^2)} \, {q_\alpha \,
q_n \over q^2} \, s({\vec q},\omega) &(5.12b)\cr
G_{0 \, {\tilde \sigma} \sigma}({\vec q},\omega) &= {1 \over - i
\omega + \lambda_0 (m_0^2 + q^2)} \! \left[ 1 - {q_n^2 \over q^2}
\, {- i \omega + \lambda_0 q^2 \over - i \omega + \lambda_0
(m_0^2 + q^2)} \, s({\vec q},\omega) \right] ; \quad
&(5.12c)\cr}$$
here the abbreviation
$$s({\vec q},\omega) = {\lambda_0 \, g_0 \over - i \, \omega +
\lambda_0 \, (g_0 + q^2) - \lambda_0 \, g_0 \, {\lambda_0 \,
m_0^2 \over - i \, \omega + \lambda_0 \, (m_0^2 + q^2)} \, q_n^2
/ q^2} \eqno(5.12d)$$
has been introduced. Note that in the static limit $\omega
\rightarrow 0$ they acquire the form
$$G_{0 \, \alpha \beta}({\vec q}) = {1 \over r_0^\alpha + q^2} \,
\left( \delta_{\alpha \beta} - g_0 \, {q_\alpha \, q_\beta \over
r_0^\beta + q^2} \, {1 \over q^2 + g_0 \, {\tilde q}^2} \right)
\quad , \eqno(5.13a)$$
where
$${\tilde q}^2 = \sum_\alpha {q_\alpha^2 \over r_0^\alpha + q^2}
\quad , \eqno(5.13b)$$
which is precisely the result quite generally derived by Aharony
[24] for anisotropic exchange accompanied by dipolar
interactions, if $r_0^\alpha = 0$ for $\alpha = 1 , \ldots , n-1$
and $r_0^n = m_0^2$ is inserted.

The rather complex structure of the result (5.12) for the
propagators of the harmonic theory renders a complete treatment
of the crossover regime a very streneous task. Hence we shall
restrict our discussion to the coexistence limit $m_0 \rightarrow
\infty$ only, which suffices for the questions we have in mind.
Before embarking on that, however, we are going to discuss very
briefly the situation at the critical point.

With $m_0 = 0$, the response and correlation propagators at $T_c$
read
$$\eqalignno{G_{0 \, {\tilde \phi}^\alpha \phi^\beta}({\vec
q},\omega) &= {1 \over - i \, \omega + \lambda_0 \, q^2} \,
P^T_{\alpha \beta} + {1 \over - i \, \omega + \lambda_0 \, (g_0 +
q^2)} \, P^L_{\alpha \beta} &(5.14a)\cr
G_{0 \, \phi^\alpha \phi^\beta}({\vec q},\omega) &= {2 \,
\lambda_0 \over \omega^2 + \lambda_0^2 \, q^4} \, P^T_{\alpha
\beta} + {2 \, \lambda_0 \over \omega^2 + \lambda_0^2 \, (g_0 +
q^2)^2} \, P^L_{\alpha \beta} \quad . &(5.14b)\cr}$$
For $T \not= T_c$, one may reintroduce the temperature dependence
$r_0$ (see Ref.[37]). The $T_c$-shift can be implicitly
determined to first order in $u_0$, e.g. by using the equation of
state (2.11), as the solution of the equation
$$g_0 = r_{0c} \left[ \left( {12 \, \epsilon \over u_0 \, A_d} \,
r_{0c}^{\epsilon / 2} - 2 \, n - 3 \right)^{2 / (2 - \epsilon)} -
1 \right] \quad . \eqno(5.15)$$
Compared to the isotropic case, the transition temperature is
increased due to the long-range order dipol-dipol interaction
which favors condensation [31].

For the renormalization in the critical region conceptually
different $Z$-factors for the transverse and longitudinal field
components, respectively, have to be defined [37], namely
$$\eqalignno{{\tilde \phi}^\alpha &= Z_{\tilde T}^{1/2} \,
\sum_\beta P^T_{\alpha \beta} \, {\tilde \phi}_0^\beta +
Z_{\tilde L}^{1/2} \, \sum_\beta P^L_{\alpha \beta} \, {\tilde
\phi}_0^\beta &(5.16a)\cr
\phi^\alpha &= Z_T^{1/2} \, \sum_\beta P^T_{\alpha \beta} \,
\phi_0^\beta + Z_L^{1/2} \, \sum_\beta P^L_{\alpha \beta} \,
\phi_0^\beta \quad . &(5.16b)\cr}$$
The renormalized counterparts for the timescale $\lambda_0$ and
the coupling constants $u_0$, $g_0$ are conventionally introduced
as
$$\eqalignno{\lambda &= Z_\lambda^{-1} \, \lambda_0 &(5.17a)\cr
u &= Z_u^{-1} \, u_0 \, A_d \, \mu^{- \epsilon} &(5.17b)\cr
g &= Z_g^{-1} \, g_0 \, \mu^{-2} \quad . &(5.17c)\cr}$$
As a consequence of the fluctuation-dissipation theorem (2.8) and
the partition of the propagators into transverse and longitudinal
components according to (5.14) the following relation between
$Z_\lambda$ and the field renormalizations can be derived
$$Z_\lambda = {Z_T^{-1} + Z_L^{-1} \over (Z_{\tilde T} \, Z_T)^{-
1/2} + (Z_{\tilde L} \, Z_L)^{-1/2}} \quad , \eqno(5.18)$$
which for $Z_{\tilde T} = Z_{\tilde L}$ and $Z_T = Z_L$ (valid
for vanishing $g_0$) reduces to the well-known expression for the
isotropic relaxational models [12,11].

In the vicinity of $T_c$, interesting crossover phenomena occur
which are induced by the second lengthscale $1/\sqrt{g_0}$
originating from the dipolar term. An elegant description of
these phenomena may be obtained by applying Amit and
Goldschmidt's general method [14]; hereby it appears that there
are no ultraviolet divergences $\propto g_0$, and hence one
concludes [37]
$$Z_g^{-1} = Z_L \quad . \eqno(5.19)$$

To one-loop order the field renormalizations vanish
$$Z_{\tilde T} = Z_{\tilde L} = Z_T = Z_L = 1 \quad ,
\eqno(5.20a)$$
and therefore
$$\eqalignno{Z_\lambda &= 1 &(5.20b)\cr
Z_g &= 1 \quad . &(5.20c)\cr}$$
For the single non-trivial $Z$-factor one finds by investigation
of the four-point vertex function [37]
$$Z_u = 1 + {n^2 + 18 \, n + 116 \over 144 \, \epsilon} \, u_0 \,
A_d \, \mu^{- \epsilon} - {n^2 - 6 \, n - 76 \over 144 \,
\epsilon} \, {u_0 \, A_d \, \mu^{- \epsilon} \over (1 + g_0 /
\mu^2)^{\epsilon / 2}} \eqno(5.20d)$$
(here we have slightly modified the results from Ref.[37] by not
explicitly inserting $n = 4$ and refraining from the
$\epsilon$-expansion).

Taking advantage of the renormalization group equation we
introduce flow-depen- dent couplings according to
$$\eqalignno{\ell \, {d g(\ell) \over d \ell} &= g(\ell) \,
\zeta_g(\ell) &(5.21a)\cr
\ell \, {d u(\ell) \over d \ell} &= \beta_u(\ell) \quad ,
&(5.21b)\cr}$$
with the usual initial conditions $g(1) = g$ and $u(1) = u$,
where
$$\zeta_g = \mu \, {\partial \over \partial \mu} \bigg \vert_0
\ln {g^2 \over g_0^2} = - 2 \eqno(5.22a)$$
and
$$\beta_u = \mu \, {\partial \over \partial \mu} \bigg \vert_0 u
= u \, \left[ - \epsilon + {n^2 + 18 \, n + 116 \over 144} \, u -
{n^2 - 6 \, n - 76 \over 144} \, {u \over (1 + g)^{1 + \epsilon /
2}} \right] \eqno(5.22b)$$
are Wilson's flow functions to one-loop order. For vanishing
dipolar interaction $g = 0$ one rediscovers the isotropic
Heisenberg fixed point (2.20a), as the zero of (5.22b). For any
finite value of $g(1)$, however, the dipolar strength diverges as
$$g(\ell) = g \, \ell^{-2} \quad , \eqno(5.23)$$
which ultimately justifies the corresponding term in (5.20d) and
(5.22b). Near the critical point where the correlation length
becomes macroscopic, the dipol-dipol interaction hence is a
relevant perturbation. The asymptotic behavior is then no more
governed by the Heisenberg fixed point but by the dipolar fixed
point [31]
$$u_D^* = {144 \, \epsilon \over n^2 + 18 \, n + 116} \quad .
\eqno(5.24)$$
The ensuing crossover features have been intensively studied by
several authors [32-37]. For the possible values $n = 2,3,4$ one
finds $u_D^* > u_H^*$, which accounts for the fact that a certain
amount of the fluctuations is effectively suppressed by the
dipolar forces.
\bigskip

\centerline{\bf B. Renormalization of the theory in the
coexistence limit for $n = d \geq 3$}
\medskip
We shall now come to the properties within the ordered phase,
where, as we shall see shortly, the cases $n = 2$ and $n = 3,4$
have to be distinguished. For $n \geq 3$ the asymptotic theory
can be discussed in a manner completely analogous to Section
II.B. Substituting the transformed longitudinal fields (2.13)
(with $a = 0$) into (5.6) and the interaction terms, and
performing the limit $m_0 \rightarrow \infty$ one finds for the
Janssen-De Dominicis functional
$$\eqalignno{J_\infty[&\{ {\tilde \pi}_0^\alpha \} , {\tilde
\sigma}_0 , \{ \pi_0^\alpha \} , \sigma_0] = &(5.25)\cr
&= \int_q \int_\omega \biggl[ \sum_{\alpha} \left( \lambda_0 \,
{\tilde \pi}_0^\alpha({\vec q},\omega) \, {\tilde \pi}_0^\alpha(-
{\vec q},-\omega) - {\tilde \pi}_0^\alpha({\vec q},\omega) \,
\Bigl[ i \, \omega + \lambda_0 \, q^2 \Bigr] \, \pi_0^\alpha(-
{\vec q},- \omega) \right) \cr
&\qquad \quad - \lambda_0 \, g_0 \, \sum_{\alpha \beta} {q_\alpha
q_\beta \over q^2} \, {\tilde \pi}_0^\alpha({\vec q},\omega) \,
\pi_0^\beta(-{\vec q},-\omega) - \lambda_0 \, {\tilde
\varphi}_0({\vec q},\omega) \, \varphi_0(-{\vec q},-\omega)
\biggr] \quad , \cr}$$
i.e.: remarkably a harmonic theory again, which can be treated
exactly.

One immediate consequence of (5.25) are the relations $\Gamma_{0
\, {\tilde \pi}^\alpha {\tilde \pi}^\beta}^\infty({\vec
q},\omega) = 2 \, \lambda_0 \, \delta_{\alpha \beta}$ and
$\Gamma_{0 \, {\tilde \pi}^\alpha \pi^\beta}^\infty({\vec
q},\omega) = B_{\alpha \beta}$ [see Eq.(5.8a)]. Hence the field
renormalizations for the $\pi$-fluctuations vanish; a similar
reasoning applies for the longitudinal fields, and from (5.18)
and (5.19) one concludes that Eqs.(5.20) are valid as exact
results in the coexistence limit. In particular, for $\ell
\rightarrow 0$ the parameter $g(\ell)$ diverges as in (5.23), and
we simply have to insert $s_\infty({\vec q},\omega) = q^2 / (q^2
- q_n^2)$ into the $\pi$-propagators:
$$\eqalignno{G_{0 \, {\tilde \pi}^\alpha \pi^\beta}^\infty({\vec
q},\omega) &= {1 \over - i \, \omega + \lambda_0 \, q^2} \,
{\overline P}^T_{\alpha \beta} &(5.26a)\cr
G_{0 \, {\tilde \pi}^\alpha {\tilde \pi}^\beta}^\infty({\vec
q},\omega) &= {2 \, \lambda \over \omega^2 + \lambda_0^2 \, q^4}
\, {\overline P}^T_{\alpha \beta} \quad . &(5.26b)\cr}$$
Note that of the two proper projectors [compare Eq.(5.3)]
$$\eqalignno{{\overline P}^T_{\alpha \beta} &= \delta_{\alpha
\beta} - {q_\alpha \, q_\beta \over q^2 - q_n^2} &(5.27a)\cr
{\overline P}^L_{\alpha \beta} &= {q_\alpha \, q_\beta \over q^2
- q_n^2} &(5.27b)\cr}$$
only the transversal operator appears in (5.26). (In the case
${\vec q} = (0,\ldots,0,q_n)$ one has ${\overline P}_{\alpha
\beta}^T = \delta_{\alpha \beta}$ and ${\overline P}_{\alpha
\beta}^L = 0$.) The fluctuations parallel to the wavevector
${\vec q}$ have disappeared, as well as the non-diagonal
propagator (5.12b).

As for the isotropic model, using (2.13), (5.25) and (5.26) one
finds the following asymptotically exact results for the
$\sigma$-correlation functions
$$\eqalignno{G_{0 \, {\tilde \sigma} \sigma}^\infty({\vec
q},\omega) = {1 \over \lambda_0 m_0^2} \Biggl( 1 &+ {u_0 \over 6}
\int_k {1 \over \left( {{\vec q} \over 2} - {\vec k} \right)^2}
\, {1 \over {- i \, \omega \over 2 \, \lambda_0} + {q^2 \over 4}
+ k^2} \times &(5.28a)\cr
&\times \left[ n - 2 - \, {\left( q^2 - q_n^2 \right) \, \left(
k^2 - k_n^2 \right) - \left( {\vec q} \, {\vec k} - q_n \, k_n
\right)^2 \over \left( {q^2 - q_n^2 \over 4} + k^2 - k_n^2
\right)^2 - \left( {\vec q} \, {\vec k} - q_n \, k_n \right)^2}
\right] \Biggr) \cr}$$
$$\eqalignno{G_{0 \, \sigma \sigma}^\infty({\vec q},\omega) = {1
\over \lambda_0 \, m_0^2} \, {u_0 \over 6} \, {\rm Re} &\int_k {1
\over \left( {q^2 \over 4} + k^2 \right)^2 - \left( {\vec q} \,
{\vec k} \right)^2} \, {1 \over {- i \, \omega \over 2 \,
\lambda_0} + {q^2 \over 4} + k^2} \times &(5.28b)\cr
&\times \left[ n - 2 - \, {\left( q^2 - q_n^2 \right) \, \left(
k^2 - k_n^2 \right) - \left( {\vec q} \, {\vec k} - q_n \, k_n
\right)^2 \over \left( {q^2 - q_n^2 \over 4} + k^2 - k_n^2
\right)^2 - \left( {\vec q} \, {\vec k} - q_n \, k_n \right)^2}
\right] \; . \cr}$$
Again, these expressions can be identified with the explicit
results from first-order perturbation theory for the two-point
cumulants in the coexistence limit. As for $g_0 = 0$, the
corresponding two-point vertex functions may be represented by a
geometric series of the $\pi$-loops, see Fig.2 of Ref.[11]. The
essential differences between (5.28) and Eq.(3.18) of Ref.[11]
are (i) the dependence on the angle between spontaneous order
parameter and wavevector as apparent from the second term in the
integrands of (5.28), and (ii) the factor $n - 2$ within the
isotropic term instead of $n - 1$.

For the determination of the only non-trivial renormalization
constant $Z_m^\infty = Z_u^\infty$ one has to consider $\Gamma_{0
\, {\tilde \sigma} \sigma}^\infty(0,0)$. Hence $Z_m$ is
independent of the anisotropies in (5.28), and one finds
explicitly that as a consequence of the dipolar interactions the
relevant effective number of massless modes is reduced from $n -
1$ to $n - 2$, because
$$Z_m^\infty = Z_u^\infty = 1 + {n - 2 \over 6 \, \epsilon} \,
u_0 \, A_d \, \mu^{- \epsilon} \quad . \eqno(5.29)$$

The remaining non-zero flow functions read
$$\zeta_g^\infty = - 2 \eqno(5.30a)$$
$$\zeta_m^\infty = - 2 + {n - 2 \over 6} \, u \eqno(5.30b)$$
$$\beta_u^\infty = u \, \left[ - \epsilon + {n - 2 \over 6} \, u
\right] \quad , \eqno(5.30c)$$
and as a solution of $\beta_u^\infty(u^*) = 0$ we find the stable
dipolar coexistence fixed point
$$u^*_{CD} = {6 \, \epsilon \over n - 2} \eqno(5.31a)$$
replacing (2.21a); however, the anomalous dimension of the mass
parameter is not altered
$$\zeta^*_{m \, CD} = - 2 + \epsilon \quad . \eqno(5.31b)$$
We remark that as a consequence of $Z_m = Z_u$ this is a general
result for every non-trivial fixed point value $u^* \not= 0$.
Therefore the power laws characteristic of the coexistence
anomalies are not changed, either. For example, the dynamic
susceptibility will have the asymptotic form
$$\chi_L({\vec q},\omega) \propto {\bar \phi}^2 \, \Bigg \vert
{q^2 \over \mu^2} - {i \, \omega \over \lambda \, \mu^2} \Bigg
\vert^{- \epsilon / 2} \quad , \eqno(5.32)$$
leading to the anomalous lineshape of the longitudinal
correlation function and the ${\vec q}$- and $\omega$-divergences
summarized in the center column of Table I. But the amplitudes of
the scaling functions (5.28) are reduced with respect to the
isotropic case, showing certain characteristic angle dependences.

In two special cases, namely if the wavevector is either parallel
or perpendicular to the spontaneous order parameter, the ${\vec
q}$-dependent contribution in the bracket of Eqs.(5.28a,b)
vanishes, and the expressions (5.28) may be directly compared
with the isotropic correlation functions for $a = 0$. Due to the
renormalization-group invariant (2.22), asymptotically $m(\ell)^2
= m(1)^2 \, (u^* / u(1)) \, \ell^{-2 + \epsilon}$, and hence the
amplitude of the longitudinal susceptibility is proportional to
$1 / u^*$. Inserting the fixed point values (5.31a) and (2.21a),
respectively, we therefore we find the following exact amplitude
ratio of the longitudinal response functions in the dipolar and
isotropic case
$${\chi_L({\vec q},\omega)_{dipolar} \over \chi_L({\vec
q},\omega)_{isotropic}} = {n - 2 \over n - 1} \quad .
\eqno(5.33)$$
For $n = 3$ this universal amplitude ratio is $1/2$, in accord
with the results of Pokrovsky [38], and Toh and Gehring [39],
obtained in the framework of a spin-wave theory. If the
wavevector ${\vec q}$ is neither parallel nor perpendicular to
the magnetization, the amplitude of (5.32) will be further
reduced.

For $n = d = 4$ there will be logarithmic corrections instead of
power-law singularities. As is apparent from (5.29)-(5.33), the
situation at $n = d = 2$ requires a separate discussion.
Afterwards an immediate generalization to the general case $n
\not= d$ will be possible.
\bigskip

\centerline{\bf C. The cases $n = d = 2$ and $n \not= d$}
\medskip
The form of the dipolar coexistence fixed point (5.31a), in
comparison with (2.21a), suggests that the effective number of
critical fluctuations is reduced to $n - 2$. Hence for $n = 2$
there are no more massless modes left that may lead to infrared
singularities. Indeed, the transverse projector (5.27a)
disappears in this case, and inserting $n = 2$ into (5.30) one
arrives at the following asymptotic behavior for the
$\ell$-dependent parameters
$$\eqalignno{g(\ell) &\propto \ell^{-2} &(5.34a)\cr
m(\ell) &\propto \ell^{-1} &(5.34b)\cr
u(\ell) &\propto \ell^{- \epsilon} \quad . &(5.34c)\cr}$$
According to the preceding investigations in Section III and
Section IV.B we may readily identify this as the signature of a
crossover to a Gaussian theory. The fluctuations die out on
leaving the critical region $T \approx T_c$, the $\sigma$-mode
because of the formation of long-range order, and the
$\pi$-excitation due to the dipolar coupling. The above
discussion of the asymptotic theory thus does not apply for the
two-component model, and the effective exponents characterizing
the static susceptibilities, for example, will rather look
similar to Fig.2.

Remarkably, there appears no inconsistency for the assumption of
long-range order in two dimensions, as opposed to the isotropic
model. This, again, is no contradiction to the
Mermin-Wagner-Hohenberg theorem [22,23] which is based on the
assumption of solely short-range isotropic interactions. We would
like to emphasize once again that more complicated situations, as
inhomogeneous ordering or topological excitations, are not within
the scope of the present formalism.

We shall now drop the restriction (5.5) and return to the general
situation where the number of components $n$ and the spatial
dimension $d$ may be different. Let us begin with the case $d >
n$; the upper limits of the sums in (5.1) and (5.4) are then $n$
and the second term in (5.4) is absent. Hence the structure of
the asymptotic Janssen-De Dominicis functional (5.25) is not
altered. We only have to be careful when inverting matrix B of
Eq.(5.8a), resulting in
$$B^{-1}_{\alpha \beta}({\vec q},\omega) = {1 \over i \, \omega +
\lambda_0 \, q^2} \left( \delta_{\alpha \beta} - {\lambda_0 \,
g_0 \over i \, \omega + \lambda_0 \, (g_0 + q^2) - \lambda_0 \,
g_0 \, \left( 1 - Q^2 / q^2 \right)} \, {q_\alpha \,
q_\beta \over q^2} \right) \eqno(5.35a)$$
instead of (5.9); here we have used the notation
$${\vec Q} = (q_1,\ldots,q_{n-1}) \quad . \eqno(5.35b)$$
In the coexistence limit the transverse propagators read
$$\eqalignno{G_{0 \, {\tilde \pi}^\alpha \pi^\beta}^\infty({\vec
q},\omega) &= {1 \over - i \, \omega + \lambda_0 \, q^2} \,
\left( \delta_{\alpha \beta} - \, {q_\alpha \, q_\beta \over Q^2}
\right) &(5.36a)\cr
G_{0 \, {\tilde \pi}^\alpha {\tilde \pi}^\beta}^\infty({\vec
q},\omega) &= {2 \, \lambda \over \omega^2 + \lambda_0^2 \, q^4}
\, \left( \delta_{\alpha \beta} - \, {q_\alpha \, q_\beta \over
Q^2} \right) \quad , &(5.36b)\cr}$$
and for the longitudinal correlation functions one arrives at
$$\eqalignno{G_{0 \, {\tilde \sigma} \sigma}^\infty({\vec
q},\omega) = {1 \over \lambda_0 m_0^2} \Biggl( 1 + {u_0 \over 6}
&\int_k {1 \over \left( {{\vec q} \over 2} - {\vec k} \right)^2}
\, {1 \over {- i \, \omega \over 2 \, \lambda_0} + {q^2 \over 4}
+ k^2} \times &(5.37a)\cr
&\qquad \times \left[ n - 2 - \, {Q^2 \, K^2 - \left( {\vec Q} \,
{\vec K} \right)^2 \over \left( {Q^2 \over 4} + K^2 \right)^2 -
\left( {\vec Q} \, {\vec K} \right)^2} \right] \Biggr) \cr}$$
$$\eqalignno{G_{0 \, \sigma \sigma}^\infty({\vec q},\omega) = {1
\over \lambda_0 \, m_0^2} \, {u_0 \over 6} \, {\rm Re} &\int_k {1
\over \left( {q^2 \over 4} + k^2 \right)^2 - \left( {\vec q} \,
{\vec k} \right)^2} \, {1 \over {- i \, \omega \over 2 \,
\lambda_0} + {q^2 \over 4} + k^2} \times &(5.37b)\cr
&\qquad \times \left[ n - 2 - \, {Q^2 \, K^2 - \left( {\vec Q} \,
{\vec K} \right)^2 \over \left( {Q^2 \over 4} + K^2 \right)^2 -
\left( {\vec Q} \, {\vec K} \right)^2} \right] \quad , \cr}$$
which obviously generalizes Eq.(5.28).

In particular the renormalization constants are identical to the
previous results; therefore for $d = 4$ and $n = 3$ there will be
logarithmic corrections characteristic for the upper critical
dimension, and for $d = 3 , 4$ and $n = 2$ a crossover to a
Gaussian behavior will be found. Thus for $d > n$ there are no
specifically interesting modifications with respect to the cases
under discussion before.

In the opposite situation, $n > d$, it is crucial whether the
spontaneous order is created in one of the directions that are
affected by the dipolar interaction, or not. One would expect the
former situation to apply, because the dipolar forces favor
condensation. Yet if there are further anisotropies present, even
transitions between different orientations e.g. of the
magnetization of two-dimensional Heisenberg ferromagnets may
occur [40]; consequently we shall take both possibilities into
account. The modified matrix $B^{-1}_{\alpha \beta}$ keeps the
form (5.35a), albeit with
$${\vec Q} = (q_{n-d+1},\ldots,q_{n-1}) \quad , \eqno(5.38a)$$
if the $\sigma$-modes are subject to the dipol-dipol forces, and
$${\vec Q} = (q_1,\ldots,q_d) \quad , \eqno(5.38b)$$
if they are not. In the first case there will be $d - 1$
transverse propagators of the type (5.36) with $\alpha , \beta =
n-d+1 , \ldots , n-1$, and $n-d$ propagators
$$\eqalignno{G_{0 \, {\tilde \pi}^\alpha \pi^\beta}^\infty({\vec
q},\omega) &= {1 \over - i \, \omega + \lambda_0 \, q^2} \,
\delta_{\alpha \beta} &(5.39a)\cr
G_{0 \, {\tilde \pi}^\alpha {\tilde \pi}^\beta}^\infty({\vec
q},\omega) &= {2 \, \lambda \over \omega^2 + \lambda_0^2 \, q^4}
\, \delta_{\alpha \beta} &(5.39b)\cr}$$
with $\alpha , \beta = 1 , \ldots , n-d$ that are equivalent to
those of the isotropic model. If the $\sigma$-field is not
affected by the dipolar interaction, the first $d$ transverse
modes have the form (5.36), while the remaining $n-d-1$
propagators are given by Eq.(5.39).

The $\sigma$-correlations and hence the longitudinal (with
respect to the order parameter) response functions look like
(5.37), where one has to insert (5.38a) or (5.38b), respectively.
The relevant asymptotic $Z$-factor is again (5.29), and for
$\epsilon > 0$ the dipolar coexistence fixed point determines the
emerging coexistence anomalies. The leading term (in the limit
${\vec q} \rightarrow 0$ and $\omega \rightarrow 0$) of the
longitudinal susceptibility is given by (5.32), accompanied by an
angle-dependent scaling function which assumes a maximum if
${\vec Q} = 0$, and is otherwise smaller than unity.

Thus the results of Subsection B can be taken over quite easily
to the general case with arbitrary number of components $n \geq
2$ and dimension $2 < d \leq 4$. The qualitative features are
summarized in Table III.
\bigskip \bigskip

\centerline{\bf VI. SUMMARY AND DISCUSSION}
\bigskip
For isotropic models, $n-1$ massless Goldstone modes appear in
the phase with spontaneously broken $O(n)$-symmetry. Their
origin is that no special direction of the spontaneous order
parameter is preferred by the static Hamiltonian (2.1); hence an
infinitesimal rotation of ${\bar \phi}$ costs no (free) energy.
These gapless excitations induce so-called coexistence anomalies
in the entire ordered phase, see Table I [11]. Within our
formalism [11], which is based on the crossover theory of Amit
and Goldschmidt [14] and generalizes Lawrie's work [8] to
dynamical problems, their analytical form can be traced back to
the renormalization group fixed point $u_C^*$ (2.21a), or rather
the accompanying anomalous dimension $\zeta_{m \, C}^*$ (2.21b).
Remarkably, the coexistence limit ($T < T_c$, ${\tilde h} = 0$,
${\vec q} \rightarrow 0$, $\omega \rightarrow 0$) can be exactly
represented by a one-loop theory for the two-point cumulants or
the leading order (spherical limit) of the $1/n$-expansion for
the two-point vertex functions.

For the application to real systems, in particular to solids
displaying structural or magnetic phase transitions, quite
naturally the question arises whether certain anisotro- pic terms
in the free-energy functional will completely destroy the
coexistence anomalies, or if, and then under which circumstances,
remnants of the Goldstone excitations will still affect the
static and dynamic properties. In order to elucidate this issue,
we have studied two characteristic cases, namely (i) cubic
anharmonicity, and (ii) dipol-dipol interaction. Generically,
three possible situations are conceivable: (a) the anisotropy
immediately suppresses the transverse fluctuations, as will be
the case for uniaxial dipolar forces or any mechanism directly
introducing transverse mass terms, and also for the cubic terms
provided $n > n_c$. However, (b) there may also be a
quasi-Goldstone behavior in an intermediate range within the
crossover from the critical to the Gaussian regime, especially if
near $T_c$ fluctuations tend to restore the $O(n)$-symmetry, an
example of which is the cubic model with $n < n_c$ [25-27]. Even
more subtle is the dipolar interaction, where (c) although the
model is no more invariant with respect to a continuous symmetry
transformation, not all the transverse modes lose their massless
character, but only their "effective" number is reduced. Hence we
believe that the specific anisotropies we have treated here,
besides their obvious importance for many real systems, supply a
qualitative picture for what may happen rather generally.

In order to show that by means of our renormalization procedure a
coherent description of the crossover from a critical theory to
an asymptotically Gaussian model may be obtained, we discussed
the time-dependent Ising model (or rather its field-theoretical
representation) in the ordered phase. Although we have only
pointed out the features of the single-component model for $T <
T_c$, it should be quite obvious how generalizations to other
situations could be achieved. The effective couplings appearing
in Wilson's flow functions vanish in the limit $\ell \rightarrow
0$, and hence the power laws are eventually characterized by
mean-field exponents. On the other hand, the expansion parameter
of the perturbation series for the scaling functions themselves
acquires a finite, but non-universal value. Of course, none of
the resulting integrals will contain any infrared divergence.
This description remains valid until additional contributions
come into play which are considered as irrelevant with respect to
the critical properties and thus have been omitted in the
Hamiltonian.

After these preliminary discussions we have treated the complete
crossover region for the relaxational models A and B with cubic
anharmonicity. As for the isotropic system, the renormalization
constants and flow functions were evaluated to one-loop order,
and the ensuing flow equations were solved numerically. An
important distinction has to be made between the cases $n \geq
n_c = 4$ (to one-loop order), where a direct transition to the
uncritical behavior takes place, and $n < n_c$, where in an
intermediate wavenumber and frequency range coexistence-type
singularities appear. The latter crossover scenario crucially
requires that in the vicinity of the critical point the
$O(n)$-symmetry is dynamically restored by fluctuation effects
[25-27]. Lawrie's coexistence fixed point becomes unstable with
respect to cubic distortions. This we have studied by using
several values for the renormalized parameter $y$ in Figs.5,7-9,
although the latter is merely a qualitative measure of the
microscopic anisotropy. If $y < 10^{-4}$, all the characteristic
features of Goldstone anomalies in three dimensions are still
present in the effective static exponent $2 - \eta_{eff}^L$
(4.23), and they are even more prominent in the frequency
dependence of the dynamic response and correlation functions.
Therefore experiments investigating the low-frequency behavior of
the dynamical susceptibility would be very enlightening: We
emphasize that any hint of a minimum of the corresponding
effective exponent (compare Figs.7-9) would be a clear signature
of the relevance of Goldstone modes. Table II provides an
overview for the relevance of coexistence anomalies in the cubic
case as the dimension $d$ and the number of components $n$ are
varied.

On the other hand, the dipol-dipol interaction induces an
anisotropy in momentum space, which also explicitly breaks the
$O(n)$-symmetry. Yet the Goldstone modes are not entirely
extinguished, but merely their number is in effect reduced by
one. Hence while for $n = 2$ a crossover to an asymptotically
uncritical theory takes place, for $n \geq 3$ coexistence
anomalies persist, governed by the dipolar fixed point (5.31a).
Unfortunately the considerably more complex structure of the
dipolar propagators (5.12) of model A renders a complete
crossover theory a rather cumbersome task. However, the
asymptotic model, characterized by diverging mass $m(\ell)
\rightarrow \infty$ and dipolar coupling $g(\ell) \rightarrow
\infty$, may be considered on a very similar basis as in the
isotropic case. Again, a one-loop theory for the two-point
cumulants and the corresponding geometric series for the
two-point vertex functions provides an exact representation in
the ordered phase for ${\vec q} \rightarrow 0$ and  $\omega
\rightarrow 0$. As a consequence of the anomalous dimension
(5.31b) remaining unaltered, the asymptotic power laws for $n
\geq 3$ are identical to those we have found for $g = 0$, see
Table I. The anisotropy appears in the ${\vec q}$-dependence of
the relevant scaling functions (5.37), in particular with respect
to the angle between order parameter and external wavevector. If
the vector ${\vec Q}$ of Eqs.(5.35b) or (5.38a,b), respectively,
is not equal to zero, the amplitudes are generally smaller than
in the isotropic case [compare (5.33)]. A qualitative summary of
the different scenarios is given in Table III.

We finally remark that all our results for the models discussed
in this paper are in accord with exact theorems, especially
concerning the possibility of long-range order in two dimensions
[22,23]. Our formalism also allows extensions to more complicated
models containing mode-coupling vertices, which usually
constitute a relevant part of the dynamics of $O(n)$-symmetric
systems. In this paper we rather intended to study the influence
of typical anisotropies than modifications by these reversible
forces. Most of the qualitative conclusions based on the
relaxational models are expected to hold also for models with
mode-coupling terms. For example, in ferromagnets the dipolar
coupling $g(\ell)$ diverges faster than the longitudinal mass
parameter $m(\ell)$ [compare Eq.(5.30)]. Hence near $T_c$ the
influence of the dipolar interaction should dominate the effects
due to the formation of a finite order parameter, as has been
anticipated by K\"otzler, Kaldis, Kamleiter and Weber in the
interpretation of their experiments on $EuS$ below $T_c$ [41]. A
detailed analysis, however, requires a treatment of model J [1]
appropriate for the dynamics of isotropic ferromagnets.
Concerning the statics of such dipolar ferromagnets, we emphasize
once again that the present theory is already complete in the
coexistence limit.
\bigskip \bigskip

\centerline{\bf ACKNOWLEDGMENTS}
\bigskip
This work has been supported by the Deutsche
Forschungsgemeinschaft (DFG) under Contract No. Schw. 348/4-1.
U.C.T. would like to thank E. Frey and H. Schinz for most useful
discussions.
\vfill \eject

\centerline{\bf APPENDIX: CUBIC TWO-POINT VERTEX FUNCTIONS}
\bigskip
In this Appendix we list the zero- and one-loop-diagrams
(Fig.10), and the corresponding analytical results for the
two-point vertex functions of the relaxational models A ($a = 0$)
and B ($a = 2$) including weak cubic anisotropy. The integration
over the internal frequency has already been performed via the
residue theorem. (Compare Appendix B of Ref.[11].)
\bigskip

$\Gamma_{0 \, {\tilde \pi} \pi}({\vec q},\omega):$
\smallskip
${\scriptstyle \quad (a) \, = \, i \, \omega + \lambda_0 \, q^a}
\left( {\scriptstyle {\overline m}_0^2 + q^2 + {2 \, b_0 + v_0
\over 6} \, (m_0^2 - {\overline m}_0^2)} \int_k {1 \over \left(
{\overline m}_0^2 + k^2 \right) \left( m_0^2 + k^2 \right) }
\right)$
\smallskip
${\scriptstyle \quad (b) \, = \, - \lambda_0 \, q^a \, {u_0^2
\over 3 \, b_0} \, m_0^2} \int_k {\left( {{\vec q} \over 2} +
{\vec k} \right)^a \over m_0^2 + \left( {{\vec q} \over 2} -
{\vec k} \right)^2} \, {1 \over {i \, \omega \over \lambda_0} +
\left( {{\vec q} \over 2} + {\vec k} \right)^a \left[ {\overline
m}_0^2 + \left( {{\vec q} \over 2} + {\vec k} \right)^2 \right] +
\left( {{\vec q} \over 2} - {\vec k} \right)^a \left[ m_0^2 +
\left( {{\vec q} \over 2} - {\vec k} \right)^2 \right]}$
\smallskip
${\scriptstyle \quad (c) \, = \, - \lambda_0 \, q^a \, {u_0^2
\over 3 \, b_0} \, m_0^2} \int_k {\left( {{\vec q} \over 2} +
{\vec k} \right)^a \over {\overline m}_0^2 + \left( {{\vec q}
\over 2} - {\vec k} \right)^2} \, {1 \over {i \, \omega \over
\lambda_0} + \left( {{\vec q} \over 2} + {\vec k} \right)^a
\left[ m_0^2 + \left( {{\vec q} \over 2} + {\vec k} \right)^2
\right] + \left( {{\vec q} \over 2} - {\vec k} \right)^a \left[
{\overline m}_0^2 + \left( {{\vec q} \over 2} - {\vec k}
\right)^2 \right]}$
\bigskip

$\Gamma_{0 \, {\tilde \sigma} \sigma}({\vec q},\omega):$
\smallskip
${\scriptstyle \quad (d) \, = \, i \, \omega + \lambda_0 \, q^a
\left( m_0^2 + q^2 \right)}$
\smallskip
${\scriptstyle \quad (e) \, = \, - \lambda_0 q^a {n - 1 \over 3}
{u_0^2 \over b_0} m_0^2} \int_k \! {\left( {{\vec q} \over 2} +
{\vec k} \right)^a \over {\overline m}_0^2 + \left( {{\vec q}
\over 2} - {\vec k} \right)^2} {1 \over {i \omega \over
\lambda_0} + \left( {{\vec q} \over 2} + {\vec k} \right)^a
\left[ {\overline m}_0^2 + \left( {{\vec q} \over 2} + {\vec k}
\right)^2 \right] + \left( {{\vec q} \over 2} - {\vec k}
\right)^a \left[ {\overline m}_0^2 + \left( {{\vec q} \over 2} -
{\vec k} \right)^2 \right]}$
\smallskip
${\scriptstyle \quad (f) \, = \, - \lambda_0 \, q^a \, 3 \, b_0
\, m_0^2} \int_k {\left( {{\vec q} \over 2} + {\vec k} \right)^a
\over m_0^2 + \left( {{\vec q} \over 2} - {\vec k} \right)^2} \,
{1 \over {i \, \omega \over \lambda_0} + \left( {{\vec q} \over
2} + {\vec k} \right)^a \left[ m_0^2 + \left( {{\vec q} \over 2}
+ {\vec k} \right)^2 \right] + \left( {{\vec q} \over 2} - {\vec
k} \right)^a \left[ m_0^2 + \left( {{\vec q} \over 2} - {\vec k}
\right)^2 \right]}$
\bigskip

$\Gamma_{0 \, {\tilde \pi} {\tilde \pi}}({\vec q},\omega):$
\smallskip
${\scriptstyle \quad (g) \, = \, - 2 \, \lambda_0 \, q^a}$
\smallskip
${\scriptstyle \quad (h) \, = \, - 2 \lambda_0 q^{2a} {u_0^2
\over 3 \, b_0} \, m_0^2 \, {\rm Re}} \int_k {1 \over \left[
{\overline m}_0^2 + \left( {{\vec q} \over 2} + {\vec k}
\right)^2 \right] \left[ m_0^2 + \left( {{\vec q} \over 2} -
{\vec k} \right)^2 \right]} \, {\scriptstyle \times}$
\smallskip
$\qquad \qquad \qquad \qquad \qquad \qquad \qquad {\scriptstyle
\times \,} {1 \over {i \, \omega \over \lambda_0} + \left( {{\vec
q} \over 2} + {\vec k} \right)^a \left[ {\overline m}_0^2 +
\left( {{\vec q} \over 2} + {\vec k} \right)^2 \right] + \left(
{{\vec q} \over 2} - {\vec k} \right)^a \left[ m_0^2 + \left(
{{\vec q} \over 2} - {\vec k} \right)^2 \right]}$
\bigskip

$\Gamma_{0 \, {\tilde \sigma} {\tilde \sigma}}({\vec q},\omega):$
\smallskip
${\scriptstyle \quad (i) \, = \, - 2 \, \lambda_0 \, q^a}$
\smallskip
${\scriptstyle \quad (j) \, = \, - 2 \, \lambda_0 \, q^{2a} \, {n
- 1 \over 6} \, {u_0^2 \over b_0} \, m_0^2 \, {\rm Re}} \int_k {1
\over \left[ {\overline m}_0^2 + \left( {{\vec q} \over 2} +
{\vec k} \right)^2 \right] \left[ {\overline m}_0^2 + \left(
{{\vec q} \over 2} - {\vec k} \right)^2 \right]} \, {\scriptstyle
\times}$
\smallskip
$\qquad \qquad \qquad \qquad \qquad \qquad \qquad {\scriptstyle
\times \,} {1 \over {i \, \omega \over \lambda_0} + \left( {{\vec
q} \over 2} + {\vec k} \right)^a \left[ {\overline m}_0^2 +
\left( {{\vec q} \over 2} + {\vec k} \right)^2 \right] + \left(
{{\vec q} \over 2} - {\vec k} \right)^a \left[ {\overline m}_0^2
+ \left( {{\vec q} \over 2} - {\vec k} \right)^2 \right]}$
\smallskip
${\scriptstyle \quad (k) \, = \, - 2 \, \lambda_0 \, q^{2a} \, {3
\over 2} \, b_0 \, m_0^2 \, {\rm Re}} \int_k {1 \over \left[
m_0^2 + \left( {{\vec q} \over 2} + {\vec k} \right)^2 \right]
\left[ m_0^2 + \left( {{\vec q} \over 2} - {\vec k} \right)^2
\right]} \, {\scriptstyle \times}$
\smallskip
$\quad \qquad \qquad \qquad \qquad \qquad \qquad {\scriptstyle
\times \,} {1 \over {i \, \omega \over \lambda_0} + \left( {{\vec
q} \over 2} + {\vec k} \right)^a \left[ m_0^2 + \left( {{\vec q}
\over 2} + {\vec k} \right)^2 \right] + \left( {{\vec q} \over 2}
- {\vec k} \right)^a \left[ m_0^2 + \left( {{\vec q} \over 2} -
{\vec k} \right)^2 \right]}$
\vfill \eject

\centerline{\bf REFERENCES}
\bigskip
\item{[1]} P.C. Hohenberg and B.I. Halperin, Rev. Mod. Phys. {\bf
49}, 435 (1977).
\item{[2]} E. Br\'ezin and D.J. Wallace, Phys. Rev. B {\bf 7},
1967 (1973).
\item{[3]} D.J. Wallace and R.K.P. Zia, Phys. Rev. B {\bf 12},
5340 (1975).
\item{[4]} D.R. Nelson, Phys. Rev. B {\bf 13}, 2222 (1976).
\item{[5]} G.F. Mazenko, Phys. Rev. B {\bf 14}, 3933 (1976).
\item{[6]} L. Sch\"afer and H. Horner, Z. Phys. B {\bf 29}, 251
(1978).
\item{[7]} J.F. Nicoll and T.S. Chang, Phys. Rev. A {\bf 17},
2083 (1978); J.F. Nicoll, Phys. Rev. B {\bf 21}, 1124 (1980).
\item{[8]} I.D. Lawrie, J. Phys. A: Math. Gen. {\bf 14}, 2489
(1981); {\bf 18}, 1141 (1985).
\item{[9]} H.O. Heuer, J. Phys. (Paris) Colloq. C{\bf 8}, 1561
(1988); J. Phys. A {\bf 25}, 47 (1992).
\item{[10]} L. Sch\"afer, Z. Phys. B {\bf 31}, 289 (1978).
\item{[11]} U.C. T\"auber and F. Schwabl, Phys. Rev. B {\bf 46},
3337 (1992).
\item{[12]} H.K. Janssen, Z. Phys. B {\bf 23}, 377 (1976); R.
Bausch, H.K. Janssen, and H. Wagner, {\it ibid.} {\bf 24}, 113
(1976).
\item{[13]} C. De Dominicis, J. Phys. (Paris) Colloq. C {\bf 1},
247 (1976).
\item{[14]} D.J. Amit and Y.Y. Goldschmidt, Ann. Phys. (N.Y.)
{\bf 114}, 356 (1978).
\item{[15]} R. Dengler, dissertation thesis, Technische
Universit\"at M\"unchen, 1987.
\item{[16]} A. Schorgg, diploma thesis, Technische Universit\"at
M\"unchen, 1988; A. Schorgg and F. Schwabl, Phys. Rev. B {\bf
46}, 8828 (1992).
\item{[17]} A. Schorgg and F. Schwabl, Phys. Lett. A {\bf 168},
437 (1992).
\item{[18]} D.J. Wallace, J. Phys. C {\bf 6}, 1390 (1973).
\item{[19]} G. Meissner, N. Menyh\'ard, and P. Sz\'epfalusy, Z.
Phys. B {\bf 45}, 137 (1981).
\item{[20]} D.J. Amit, {\it Field Theory, the Renormalization
Group, and Critical Phenomena}, 2nd ed. (World Scientific,
Singapore, 1984).
\item{[21]} R. Schloms, dissertation thesis,
Rheinisch-Westf\"alisch Technische Hochschule Aa-chen, 1989; R.
Schloms and V. Dohm, Nucl. Phys. B {\bf 328}, 639 (1989); Phys.
Rev. B {\bf 42}, 6142 (1990).
\item{[22]} N.D. Mermin and H. Wagner, Phys. Rev. Lett. {\bf 17},
1133 (1966).
\item{[23]} P.C. Hohenberg, Phys. Rev. {\bf 158}, 383 (1967).
\item{[24]} A. Aharony, Phys. Rev. B {\bf 8}, 3349 (1973); {\bf
8}, 3358 (1973); {\bf 8}, 4270 (1973); Phys. Rev. Lett. {\bf 31},
1494 (1973).
\item{[25]} I.J. Ketley and D.J. Wallace, J. Phys. A {\bf 6},
1667 (1973).
\item{[26]} E. Br\'ezin, J.C. Le Guillou, and J. Zinn-Justin,
Phys. Rev. B {\bf 10}, 892 (1974).
\item{[27]} T. Nattermann and S. Trimper, J. Phys. A {\bf 8},
2000 (1975).
\item{[28]} I.F. Lyuksyutov and V.L. Pokrovsky, Pis'ma Zh. Eksp.
Teor. Fiz. {\bf 21}, 22 (1975) [JETP Lett. {\bf 21}, 9 (1975)];
I.F. Lyuksyutov, Phys. Lett. A {\bf 56}, 135 (1976).
\item{[29]} J. Rudnick, Phys. Rev. B {\bf 18}, 1406 (1978).
\item{[30]} H.H. Iacobson and D.J. Amit, Ann. Phys. (N.Y.) {\bf
133}, 57 (1981).
\item{[31]} M.E. Fisher and A. Aharony, Phys. Rev. Lett. {\bf
30}, 559 (1973); Phys. Rev. B {\bf 8}, 3323 (1973); A. Aharony,
{\it ibid.} {\bf 8}, 3342 (1973).
\item{[32]} A.D. Bruce and A. Aharony, Phys. Rev. B {\bf 10},
2078 (1974); {\bf 10}, 2973 (1974).
\item{[33]} T. Nattermann and S. Trimper, J. Phys. C {\bf 9},
2589 (1976).
\item{[34]} A.D. Bruce, J.M. Kosterlitz, and D.R. Nelson, J.
Phys. C {\bf 9}, 825 (1976); A.D. Bruce, {\it ibid.} {\bf 10},
419 (1977).
\item{[35]} M.A. Santos, J. Phys. C {\bf 13}, 1205 (1980).
\item{[36]} H.S. Kogon and A.D. Bruce, J. Phys. C {\bf 15}, 5729
(1982).
\item{[37]} E. Frey, dissertation thesis, Technische
Universit\"at M\"unchen, 1989; E. Frey and F. Schwabl, J. Phys.
(Paris) Colloq. C {\bf 8}, 1569 (1988); Phys. Rev. B {\bf 42},
8261 (1990); Phys. Rev. B {\bf 43}, 833 (1991).
\item{[38]} V.L. Pokrovsky, Adv. Phys. {\bf 28}, 595 (1979).
\item{[39]} H.S. Toh and G.A. Gehring, J. Phys. - Cond. Matt.
{\bf 2}, 7511 (1990).
\item{[40]} D. Pescia and V.L. Pokrovsky, Phys. Rev. Lett. {\bf
65}, 2599 (1990).
\item{[41]} J. K\"otzler, E. Kaldis, G. Kamleiter, and G. Weber,
Phys. Rev. B {\bf 43}, 11280 (1991).
\vfill \eject

\centerline{\bf TABLE I. Coexistence anomalies of the isotropic
relaxational models.}
\bigskip
\settabs\+ \qquad \qquad \qquad \qquad ${\rm Re} \,
\chi_L({\vec q},\omega)$ \qquad &Model A \qquad &Model B\cr
\+ &Model A &Model B\cr
\medskip
\+\qquad \qquad \qquad \qquad $\chi_L({\vec q},0)$
&$\propto q^{- \epsilon}$ &$\propto q^{- \epsilon}$\cr
\+\qquad \qquad \qquad \qquad ${\rm Re} \, \chi_L(0,\omega
/ q^a)$ &$\propto \omega^{- \epsilon /2}$ &$\propto (\omega /
q^2)^{- \epsilon / 2}$ \cr
\+\qquad \qquad \qquad \qquad $G_L(0,\omega / q^a)$
&$\propto \omega^{- 1 - \epsilon /2}$ &$\propto (\omega / q^2)^{-
\epsilon}$ \cr
\bigskip \bigskip \bigskip \bigskip

\centerline{\bf TABLE II. The influence of cubic anisotropy.}
\bigskip
\settabs\+$\qquad \qquad 2 \leq n \leq 4 \quad$ &via $b_C^* = {6
\, \epsilon \over n - 1}$ &Gaussian theory &$\,$ as for $v = 0$
\cr
\+$\qquad \qquad \qquad$ &$\quad d = 2$ &$\quad d = 3$ &$\quad d
= 4$ \cr
\medskip
\+$\qquad \qquad \quad n = 1 \quad \quad$ &crossover to a
&Gaussian theory, &$\,$ as for $v = 0$ \cr
\+$\qquad \qquad 2 \leq n \leq 4 \quad$ &via $b_C^* = {6 \,
\epsilon \over n - 1}$ &$\rightarrow$ anomalies &$\;$ log. corr.
\cr
\+$\qquad \qquad \quad n > 4 \quad \quad$ &directly to $0$,
&no anomalies &$\;$ log. corr. \cr
\bigskip \bigskip \bigskip \bigskip

\centerline{\bf TABLE III. The influence of dipolar
interactions.}
\bigskip
\settabs\+$\qquad \qquad n \geq 3 \qquad$ &$\quad u_{CD}^* = {6
\, \epsilon \over n - 2} \,$ &Gaussian theory &$\,$ as for $g =
0$ \cr
\+$\qquad \qquad \qquad$ &$\quad d = 2$ &$\quad d = 3$ &$\quad d
= 4$ \cr
\medskip
\+$\qquad \qquad n = 1 \qquad$ &crossover to a &Gaussian theory,
&$\,$ as for $g = 0$ \cr
\+$\qquad \qquad n = 2 \qquad$ &crossover to a &Gaussian theory,
&$\,$ no anomalies \cr
\+$\qquad \qquad n \geq 3 \qquad$ &$\quad u_{CD}^* = {6 \,
\epsilon \over n - 2} \,$ &$\rightarrow$ anomalies &$\;$ log.
corr. \cr
\vfill \eject

\centerline{\bf FIGURE CAPTIONS}
\bigskip
\item{FIG.1.} Flow of the effective coupling $u_{eff}(\ell)$ for
$n = 1$ and $\epsilon = 1$: (a) flow diagram, (b) universal
crossover.
\item{FIG.2.} Effective exponent $2 - \eta_{eff}$ of the static
susceptibility in the case $n = 1$ and $\epsilon = 1$.
\item{FIG.3.} Basic elements of the dynamical perturbation theory
for the cubic relaxational models below $T_c$.
\item{FIG.4.} Schematic flow diagrams for $m = 0$ ($T = T_c$) in
the case $n = 2$ (a) and $n = 8$ (b); $\epsilon = 1$.
\item{FIG.5.} Flow of the effective coupling $b_{eff}(\ell)$ for
several values of the anisotropy parameter $y$ ($y = 0$
corresponds to the isotropic case); $n = 2$ and $\epsilon = 1$.
\item{FIG.6.} Schematic flow diagrams for $m > 0$ ($T < T_c$) and
$v < 0$ in the case $n = 2$ (a) and $n = 8$ (b).
\item{FIG.7.} Effective exponents $2 - \eta_{eff}^T$ (a) and $2 -
\eta_{eff}^L$ (b) of the static susceptibilities for several
values of the anisotropy parameter $y$; $n = 2$ and $\epsilon =
1$.
\item{FIG.8.} Frequency dependence of ${\rm Re} \,
\chi_L(0,\omega)$ (a) and of $G_L(0,\omega)$ (b) for the cubic
model A with different values of the anisotropy parameter $y$; $n
= 2$ and $\epsilon = 1$.
\item{FIG.9.} Frequency dependence of ${\rm Re} \,
\chi_L(0,\omega / q^2)$ (a) and of $G_L(0,\omega / q^2)$ (b) for
the cubic model B with different values of the anisotropy
parameter $y$; $n = 2$ and $\epsilon = 1$.
\item{FIG.10.} Zero- and one-loop diagrams for the two-point
vertex functions of the cubic relaxational models A ($a = 0$) and
B ($a = 2$). The corresponding explicit analytical expressions
are listed in the Appendix.
\vfill \eject
\end